\documentclass[preprint,review,12pt]{elsarticle}

\usepackage{graphicx}
\usepackage{amssymb}
\usepackage{amsfonts}
\usepackage{amsmath}
\usepackage{mathrsfs}
\usepackage{subfigure}
\usepackage{xcolor}
\usepackage{multirow}
\usepackage{caption} 

\definecolor{IgorGreen}{RGB}{76,153,0}
\definecolor{IgorBlue}{RGB}{0,0,204}
\definecolor{IgorGreen1}{RGB}{0,153,77}
\definecolor{IgorPurple}{RGB}{103,0,204}
\usepackage{flushend,cuted}

\providecommand{\U}[1]{\protect\rule{.1in}{.1in}}
\journal{Physica A}

\begin{document}

\begin{frontmatter}

\title{SIS Epidemic Spreading with Correlated Heterogeneous Infection Rates}

\author[label1]{Bo Qu\corref{cor1}\fnref{label3}}
\address[label1]{Delft University of Technology}
%\address[label2]{Mekelweg 4, 2628CD, Delft, The Netherlands}

%\cortext[cor1]{I am corresponding author}

\ead{b.qu@tudelft.nl}

\author[label1]{Huiijuan Wang}
\ead{h.wang@tudelft.nl}

\begin{abstract}
The epidemic spreading has been widely studied in homogeneous cases, where each node may get infected by an infected neighbor with the same rate. However, the infection rate between a pair of nodes, which may depend on e.g. their interaction frequency, is usually heterogeneous and even correlated with their nodal degrees in the contact network. In this paper, we aim to understand how such correlated heterogeneous infection rates influence the epidemic spreading on different network topologies. Motivated by real-world datasets, we propose a correlated heterogeneous Susceptible-Infected-Susceptible (CSIS) model which assumes that the infection rate $\beta_{ij}(=\beta_{ji})$ between node $i$ and $j$ is correlated with the degree of the two end nodes: $\beta_{ij}=c(d_id_j)^\alpha$, where $\alpha$ indicates the strength of the correlation between the infection rates and nodal degrees, and $c$ is selected so that the average infection rate is $1$ in this work. In order to understand the effect of such correlation on epidemic spreading, we consider as well the corresponding uncorrected but still heterogeneous infection rate scenario as a reference, where the original correlated infection rates in our CSIS model are shuffled and reallocated to the links of the same network topology. We compare these two scenarios in the average fraction of infected nodes in the metastable state on Erd{\"o}s-R{\'e}nyi (ER) and scale-free (SF) networks with a similar average degree. Through the continuous-time simulations, we find that, when the recovery rate is small, the negative correlation is more likely to help the epidemic spread and the positive correlation prohibit the spreading; as the recovery rate increases to be larger than a critical value, the positive but not negative correlation tends to help the spreading. Our findings are further analytically proved in a wheel network (one central node connects with each of the nodes in a ring) and validated on real-world networks with correlated heterogeneous interaction frequencies. 
\end{abstract}

\begin{keyword}
SIS model; Epidemic spreading; Complex networks 
\end{keyword}

\end{frontmatter}

% \linenumbers

\section{Introduction}
\label{Intro}
Biological, social and communication systems can be represented as networks by considering the system components or individuals as nodes and the interactions or relations in between nodes as links. Viral spreading models have been used to model processes e.g.\ the propagation of information and epidemics on such networks or complex systems \cite{daley2001epidemic,pastor2014epidemic,pastor2005epidemics,pastor2001epidemic}. The Susceptible-Infected-Susceptible (SIS) model is one of the most studied models. In the SIS model, at any time $t$, the state of a node is a Bernoulli random variable, where $X_i(t)=0$ represents that node $i$ is susceptible and $X_i(t)=1$ if it is infected. Each infected node infects each of its susceptible neighbors with an infection rate $\beta$. The infected node can be recovered to be susceptible again with a recovery rate $\delta$. Both infection and recovery processes are independent Poisson processes. The ratio $\tau\triangleq\beta/\delta$ is called effective infection rate. When $\tau$ is larger than the epidemic threshold $\tau_c$, the epidemic spreads out with a nonzero fraction of infected nodes in the metastable state. The average fraction of infected nodes $y_\infty$ in the metastable state, ranging in $[0,1]$, indicates how severe the influence of the virus is: the larger the fraction $y_\infty$ is, the more severely the network is infected.

In the classic SIS model, both the infection and recovery rates are assumed homogeneous, i.e.\ the infection rates are the same for all infected-susceptible node pairs and the recovery rates are the same for all infected nodes. However, the infection rates which can be reflected from, for example, the interaction frequencies, between nodes in real-world networks are usually heterogeneous and even dependent on the properties of the nodes. For examples, the number of flights between different pairs of airports in a month are different in the airline transportation network, and the number of collaborated papers between different pairs of authors in a year vary in a co-author network \cite{barrat2004architecture,li2004statistical}. The interaction freqeuncy is found to be correlated with the nodal degrees in e.g.\ airline transportation network and metabolic network \cite{barrat2004architecture,macdonald2005minimum,li2004statistical}. Hence, we aim to understand the effect of correlated infection rate on the viral spreading in this work. We propose a correlated heterogeneous SIS (CSIS) model, in which the recovery rates are homogeneous but the infection rate $\beta_{ij} (=\beta_{ji})$ between node $i$ and $j$ is correlated with their degrees $d_i$ and $d_j$ in the way: 
\begin{equation}
\label{eq1}
\beta_{ij}=c(d_id_j)^{\alpha}
\end{equation} 
where $\alpha$ indicates the strength of the correlation and $c$ is a constant to control the average infection rate to $1$.
The correlation strength $\alpha\approx 0.5$ in the network of airports (both in US \cite{barrat2004architecture,macdonald2005minimum} and China \cite{li2004statistical}) and $\alpha\approx 0.8$ in the metabolic network \cite{macdonald2005minimum}. 

This paper aims to understand the effect of correlation between the infection rates and nodal degrees on viral spreading. Motivated by real-world networks, we consider the generic case where the heterogeneous infection rates are correlated with nodal degrees as described by (\ref{eq1}) and the network topology is as well heterogeneous. We consider also the corresponding uncorrelated heterogeneous infection rates scenario, where  the correlated infection rates are shuffled and randomly assigned to all the links as a reference scenario.
For the heterogeneity of the network topology, we consider the different broadness of the degree distribution, i.e.\ Erd{\"o}s-R{\'e}nyi (ER) and scale-free (SF) networks.
We explore how the positive or negative correlation between the infection rates and nodal degrees influences the epidemic spread on the networks with different heterogeneities of the topology.  

A few recent papers have taken into account the heterogeneous infection rates in the SIS model. We studied the SIS model with identically and independently distributed (i.i.d.) infection rates and found that independent of the underlying network topology, though the heterogeneous infection rates may increase the probability that the epidemic spreads out when the recovery rate is large, the overall infection decreases as the variance of infection rates increases if the epidemic spreads out\cite{qu2015sis}. Buono et al.\ \cite{buono2013slow} considered a specific distribution of infection rates and observed slow epidemic extinction phenomenon. Preciado et al.\ \cite{preciado2014optimal} discussed how to choose the infection and recovery rates from given discrete sets to let the virus die out. Yang and Zhou \cite{yang2012epidemic} gave an edge-based mean-field solution of the epidemic threshold in regular networks (the degrees of all nodes are the same) with i.i.d.\ heterogeneous infection rates following a uniform or power-law distribution. Fu et al.\ \cite{fu2008epidemic} analytically studied the epidemic threshold when the infection rates are as a piecewise linear function of degrees by their mean-field method. Van Mieghem and Omic \cite{van2013homogeneous} developed the N-intertwined mean field approximation (NIMFA) \cite{van2011n} for heterogeneous SIS model, and discussed the bounds of epidemic threshold and the convexity of the steady-state infection probability of each node as function of the recovery rate. Heterogeneous infection rates have also been considered in interconnected networks, allowing component networks and their interconnections to have a different but constant infection rate \cite{wang2013effect,sahneh2013generalized}. 
In this paper, we employ the continuous-time simulation of the exact SIS model instead of the mean-field methods and introduce a general correlation between the nodal degree and the infection rate which is motivated by real-world datasets to study the influence of the correlation on the epidemic spreading.

\section{Preliminary}
In this section, we first introduce the construction of network models and the heterogeneous infection rates which are considered in our CSIS model.
We then introduce the continuous-time simulation which is the main approach in this work.
Finally, we introduce the mean-field approximation of the SIS and CSIS model which will be further used in our theoretical analysis of the CSIS model on the wheel network in Section \ref{Sec_wheel}.
\subsection{Network models}
\label{sec_top}
The scale-free (SF) model has been used to capture the scale-free nature of degree distribution in real-world networks such as the Internet \cite{caldarelli2000fractal} and World Wide Web \cite{albert1999internet}: $\text{Pr}[D=d]\sim d^{-\lambda}, d\in[d_{\rm min},d_{\rm max}]$, where $d_{\rm min}$ is the smallest degree,
$d_{\rm max}$ is the degree cutoff, and $\lambda>0$ is the exponent
characterizing the broadness of the distribution
\cite{barabasi1999emergence}. In real-word networks, the exponent $\lambda$ is usually in the range $[2,3]$, thus we confine the exponent $\lambda=2.5$ in this paper. We further employ the smallest degree $d_{\rm min}=2$, the natural degree cutoff $d_{\rm max}=\lfloor N^{1/(\lambda-1)}\rfloor$ \cite{PhysRevLett.85.4626} , and the size $N=10^4$. Hence, the average degree is approximately $4$. 

The Erd\"os-R\'enyi (ER) model \cite{erdds1959random} has also been taken into account. In an ER random network with $N$ nodes, each pair of nodes is connected with a probability $p$ independent of the connection of any other pair. The distribution of the degree of a random node is binomial: $\text{Pr}[D=d]=\binom{N-1}{d}p^d(1-p)^{N-1-d}$ and the average degree is $E[D]=(N-1)p$. For large $N$ and constant $E[D]$, the degree distribution is Poissonian: $\text{Pr}[D=d]=e^{-E[D]}E[D] ^{d}/d!$. We consider the ER networks with the size $N=10^4$ and the average degree $E[D]=4$.

%The variance of nodal degrees in an ER network is much smaller than that in a SF network with the same average degree. We say that SF networks have a higher heterogeneity in topology. %Moreover, in SF networks, a smaller $\gamma$ leads to a broader distribution of degrees, thus a higher heterogeneity in topology. 

\subsection{The infection rates}
Given the network topology, we build two heterogeneous infection-rate scenarios: 1) the correlated infection rates and 2) the uncorrelated or shuffled infection rates. In the scenario of correlated infection rates, we assume that $\beta _{ij}=c\left( d_{i}d_{j}\right) ^{\alpha }$ where $\alpha$ indicates the correlation strength. We selected the constant $c$ such that the average infection rate is $1$, whereas we consider different values of homogeneous recovery rates. In this case, the infection rate of each link is determined by the given network topology and $\alpha$. In the scenario of uncorrelated infection rates, we shuffle the infection rates from all the links as generated in the first scenario and redistribute them randomly to all the links. In this way, we keep the distribution of infection rates but effectively remove the correlation between the infection rates and nodal degrees. The homogeneous infection rate is a special case of our heterogeneous infection rate construction where $\alpha=0$. Clearly, in a homogeneous network where all the nodes have the same number of neighbors, the infection rates are homogeneous in both scenarios for any $\alpha$.

\begin{figure}
\centering
\subfigure[]{
\includegraphics[scale=.28]{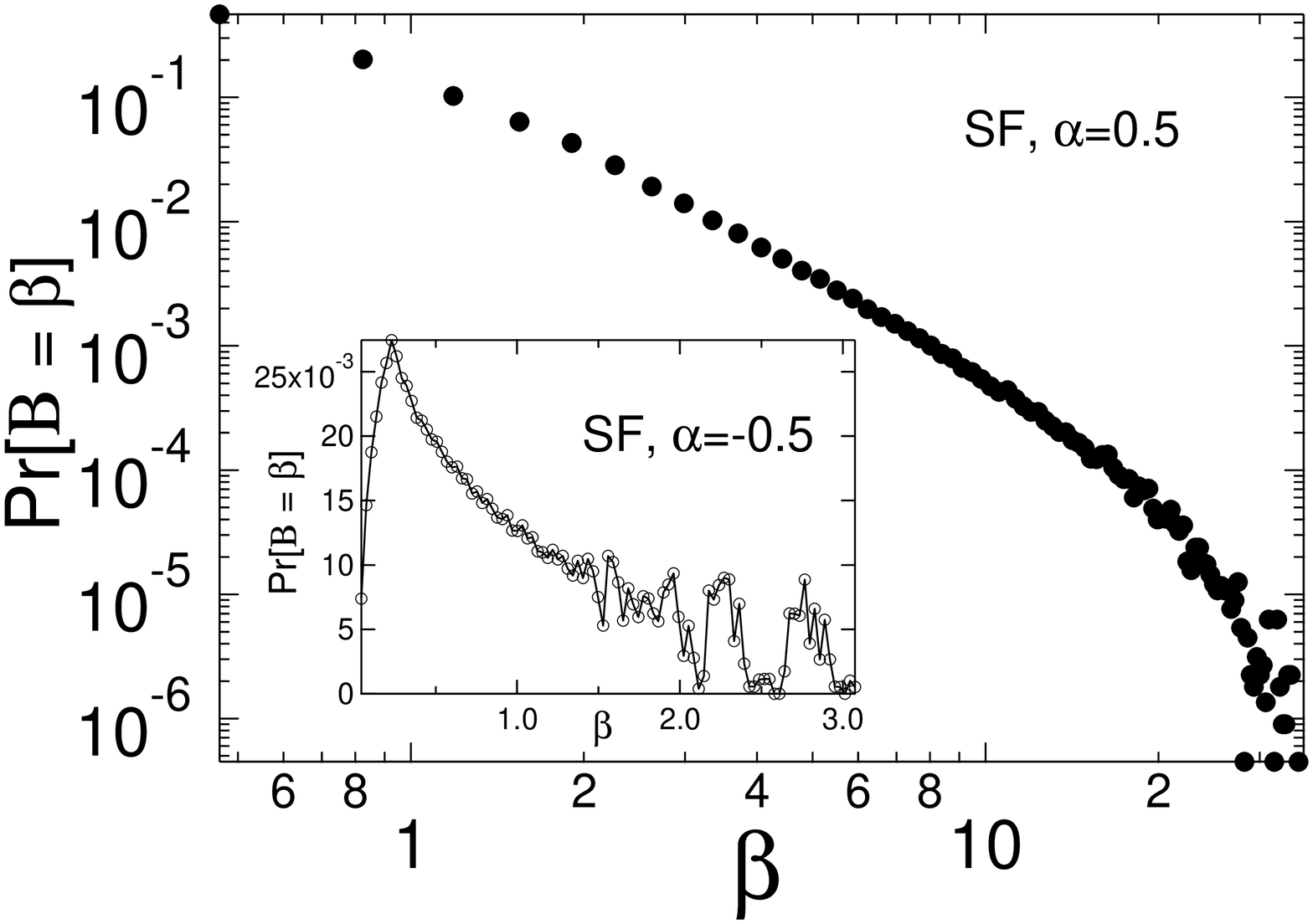}
\label{fig:01}
}
\subfigure[]{
\includegraphics[scale=.28]{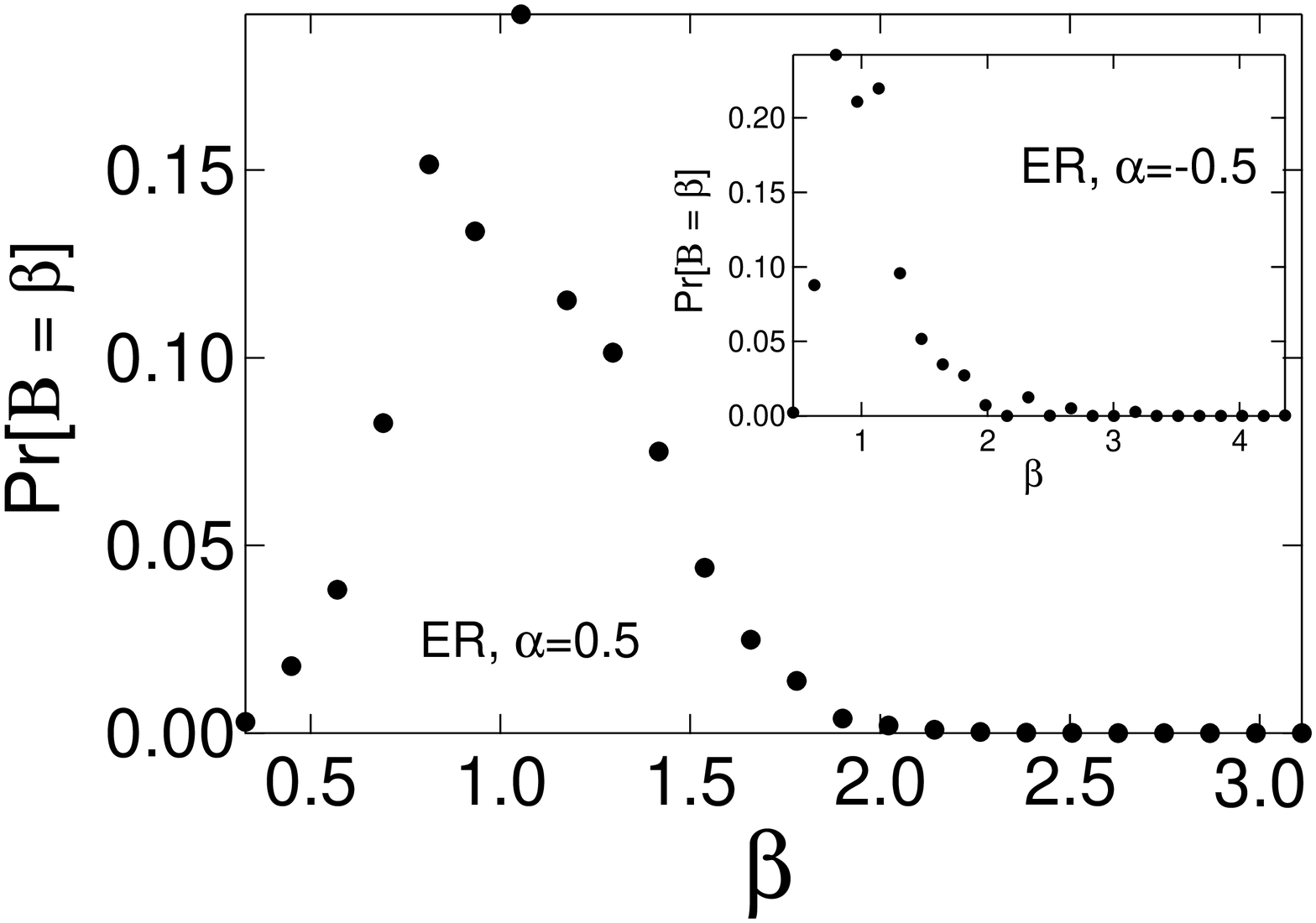}
\label{fig:02}
}
\caption{The distribution of the heterogeneous infection rates for (a) SF networks and (b) ER networks, where the parameter $\alpha=0.5$ in the main figures and $\alpha=-0.5$ in the insets. }
\label{fig:0}
\end{figure}

As examples, we show the distribution of the heterogeneous infection rates when $\alpha=0.5$ (the positive correlation) and $\alpha=-0.5$ (the negative correlation) for both ER and SF networks in Fig.~\ref{fig:0}. We find that the positive correlation between the infection rates and the nodal degrees leads to a heavy-tailed distribution of the infection rate in SF networks. However, the negative correlation in SF networks and both kinds of correlation in ER networks only generate the infection rates distributed in a small range.

A positive $\alpha>0$ (or negative $\alpha<0$), suggests a positive (or negative) correlation between the infection rates and nodal degrees. Too large or too small values of $\alpha$ could not be realistic. For example, \cite{barrat2004architecture,macdonald2005minimum,li2004statistical} suggest that $\alpha$ is around $0.5$ or $0.8$ in their datasets, and we also find $\alpha\approx0.14$ and $-0.12$ respectively in two real-world dataset as described in Section 5. Hence, we focus mainly on the range of $\alpha\in [-1,1]$ in this paper, and discuss the extreme case when the absolute value of $\alpha$ is large in Section \ref{Sec:largeA}. 

%Our previous work \cite{qu2015sis} explored the SIS model with i.i.d.\ heterogeneous infection rates and showed that the nonzero average fraction $y_\infty$ of infected nodes tends to decrease as the variance of the i.i.d.\ infection rates increases. In this paper, the uncorrelated infection rates in Scenario 2) can be considered as being independent. In this case, our previous results can be applied to derive the relationship between the average fraction $y_\infty$ of infected nodes and the parameter $\alpha$ as shown in Fig.~\ref{fig:10}: as $\alpha$ increases when $\alpha>0$ or as $\alpha$ decreases when $\alpha<0$, the variance of the infection rates increases, hence $y_\infty$ decreases.  
\subsection{The simulations}
In this paper, we perform the continuous-time simulations of the CSIS model on both ER networks and SF networks (the heterogeneity of the network topology is thus taken into account) with $N=10000$ nodes. Given a network topology, a recovery rate $\delta$ and a value of $\alpha$, we carry out $100$ iterations. In each iteration, we construct the network as described in Section \ref{sec_top}. We generate the heterogeneous infection rates as described in (\ref{eq1}) for the scenario of the correlated infection rates and shuffle them for the scenario of uncorrelated infection rates. Initially, $10\%$ of the nodes are randomly infected. Then the infection and recovery processes of SIS model are simulated until the system reaches the metastable state where the fraction of infected nodes is unchanged for a long time. The average fraction $y_\infty$ of infected nodes is obtained over $100$ iterations for both scenarios of the correlated and uncorrelated infection rates. Moreover, for simplicity, we use $y_{\infty,c}$ and $y_{\infty,u}$ to denote the average fraction of infected nodes in the scenarios of correlated and uncorrelated infection rates respectively.  

A discrete-time simulation could well approximate a continuous process if a small time bin to sample the continuous process is selected so that within each time bin, no multiple events occur. However, a heterogeneous SIS model allows different as well large infection or recovery rates, which requires even smaller time bin size and challenges the precision of a discrete-time simulation. Hence, instead of performing discrete-time simulations, we further develop the continuous-time simulator for our CSIS model, based on the one firstly proposed by van de Bovenkamp and described in detail in \cite{li2012susceptible} for homogeneous infection rates.  

\subsection{The N-Intertwined Mean-Field Approximation of the SIS model}
The N-Intertwined Mean-Field Approximation (NIMFA) is an advanced mean-field approximation of the SIS model that takes the network topology into account. The governing equation for a node $i$ in NIMFA for the classic SIS model with the homogeneous infection and recovery rates is 
\begin{equation}
\frac{\mathrm{d} v_i(t)}{\mathrm{d} t}=-\delta v_i(t)+\beta(1-v_i(t))\sum_{j=1}^{N}a_{ij}v_{j}(t)
\end{equation}
where $v_i(t)$ is the infection probability of node $i$ at time $t$, and $a_{ij}=1$ or $0$ denotes if there is a link or not between node $i$ and node $j$. In the steady state, defined by $\frac{\mathrm{d} v_{i}(t)}{\mathrm{d} t}=0$ and $\lim_{t\to \infty} v_{i}(t)=v_{i\infty}$, we obtain the infection probability of node $i$: 
\begin{equation}
\label{Equ:NIMFA-Steady}
v_{i\infty}=1-\frac{1}{1+\tau\sum_{j=1}^{N}a_{ij}v_{j\infty}}
\end{equation}
The trivial solution of (\ref{Equ:NIMFA-Steady}) is $v_{i\infty}=0$ for all nodes and indicates the absorbing state which will be reached after an unrealistically long time for a real network that is not small in size. Before reaching the absorbing state, there could exist the metastable state, indicated by the nonzero solution of (\ref{Equ:NIMFA-Steady}).

The Laurent series of the steady-state infection probability is 
\begin{equation}
\label{Equ:NIMFA-series}
v_{i\infty}=1+\sum_{m=1}^{\infty}\eta_m(i)\tau^{-m}
\end{equation}
possesses the coefficients $\eta_1(i)=-\frac{1}{d_i}$ and 
\begin{equation*}
\eta_2(i)=\frac{1}{d_i^2}(1-\sum_{j=1}^{N}\frac{a_{ij}}{d_j})
\end{equation*}
and for $m\geq2$, the coefficients obey the recursion%
\[
\eta_{m+1}\left(  i\right)  =-\frac{1}{d_{i}}\eta_{m}\left(  i\right)  \left(
1-\sum_{j=1}^{N}\frac{a_{ij}}{d_{j}}\right)  -\frac{1}{d_{i}}\sum_{k=2}%
^{m}\eta_{m+1-k}\left(  i\right)  \sum_{j=1}^{N}a_{ij}\eta_{k}\left(
j\right)
\]
When the infection rates are heterogeneous, the governing equation becomes
\begin{equation}
\frac{\mathrm{d} v_i(t)}{\mathrm{d} t}=-\delta v_i(t)+\beta_{ij}(1-v_i(t))\sum_{j=1}^{N}a_{ij}v_{j}(t)
\end{equation}
The infection probability of a node follows
 \begin{equation}
\label{Equ:NIMFA-Steady-Hetero}
v_{i\infty}=1-\frac{\delta}{\delta+\sum_{j=1}^{N}\beta_{ij}a_{ij}v_{j\infty}}
\end{equation}

\section{Effect on the average fraction $y_\infty$ of infected nodes}
\label{sec2}

As mentioned, the average fraction $y_\infty$ of infected nodes in the metastable state indicates how severe the network is infected. In this section, we explore how the average fraction $y_\infty$ of the infected nodes depends on the parameter $\alpha$ in both of the two scenarios: correlated and uncorrelated infection rates. We mainly consider the influence of the correlation between the infection rates and nodal degrees on epidemic spreading when the recovery rate varies and the absolute value of $\alpha$ is in the range $[-1,1]$. 
The difference of the influence between ER and SF networks is also discussed. Then we briefly describe the influence of the correlation in an extreme case when the absolute value of $\alpha$ is much larger. In this case, the influence of the correlation is independent from the value of the recovery rate.  

\subsection{Realistic cases: $\alpha\in[-1,1]$}
\label{sec:3.1} 
In this subsection, we first intuitively explain that how the correlation influences the epidemic spreading as the recovery rate $\delta$ and the parameter $\alpha$ vary. To support our explanations, we then define an intermediate quantity and illustrate the effect of the correlation when the recovery rate is small and large respectively.

\begin{figure}[!t]
\centering
\subfigure[]{
\includegraphics[scale=.28]{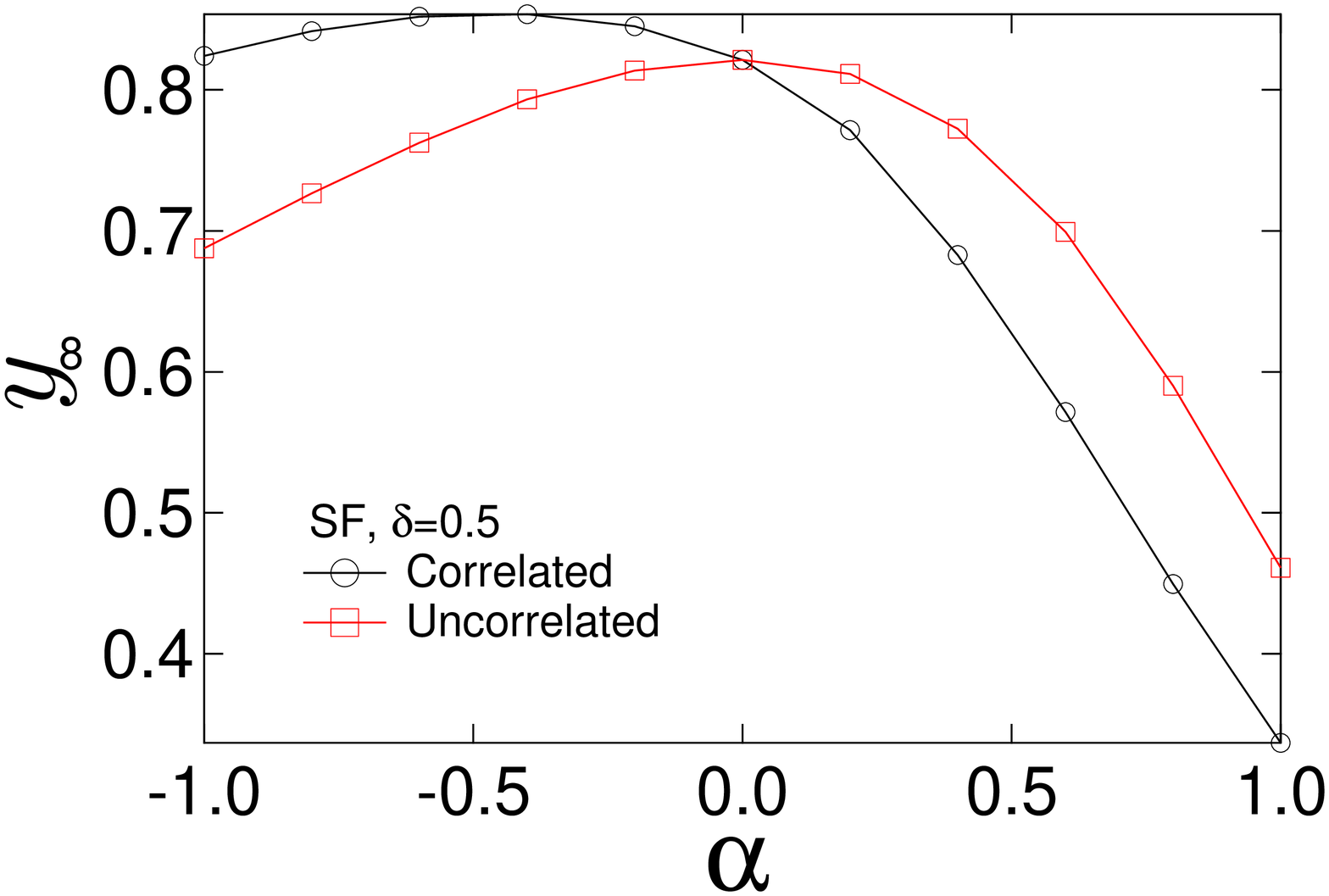}
\label{fig:11}
}
\subfigure[]{
\includegraphics[scale=.28]{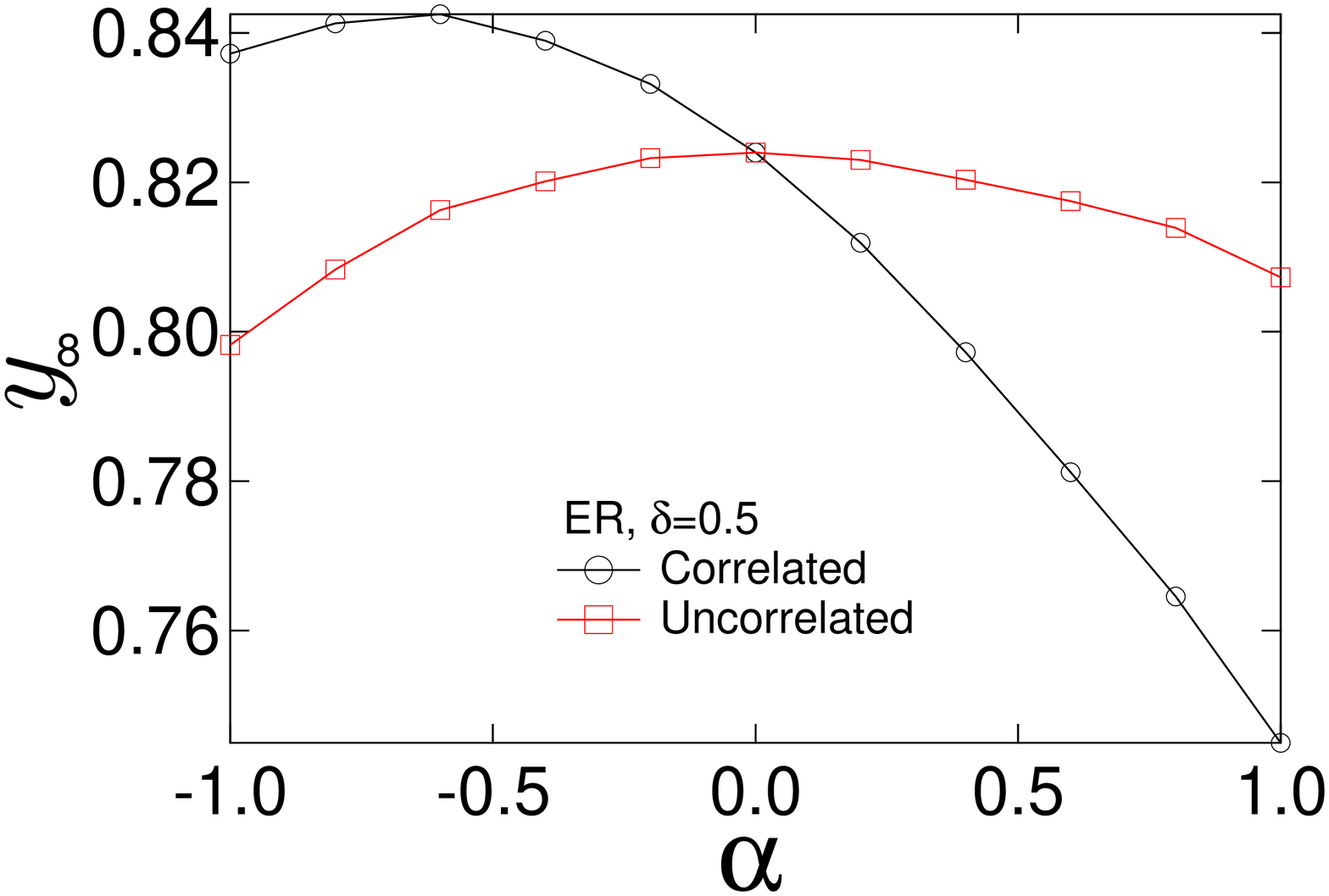}
\label{fig:12}
}
\subfigure[]{
\includegraphics[scale=.28]{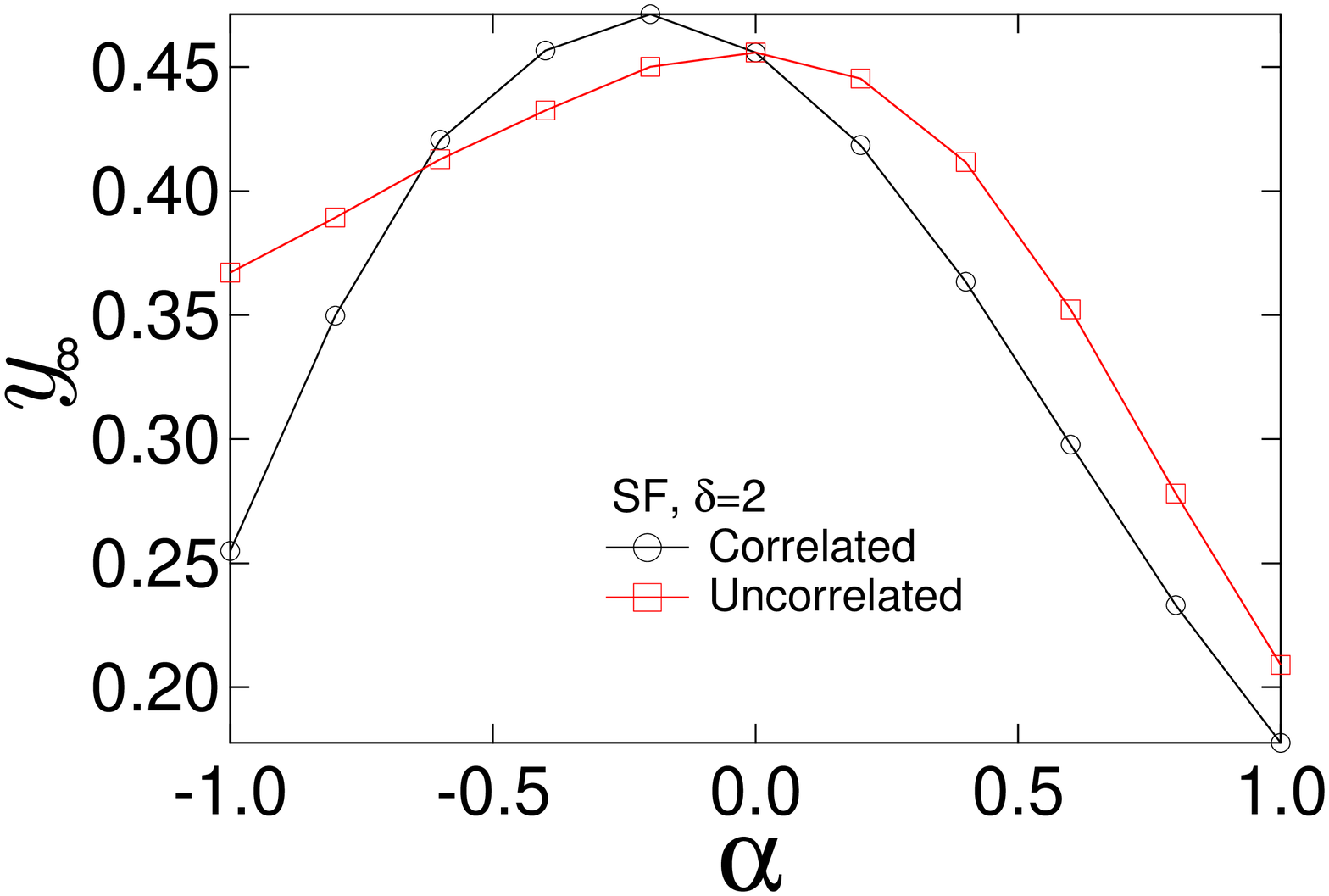}
\label{fig:13}
}
\subfigure[]{
\includegraphics[scale=.28]{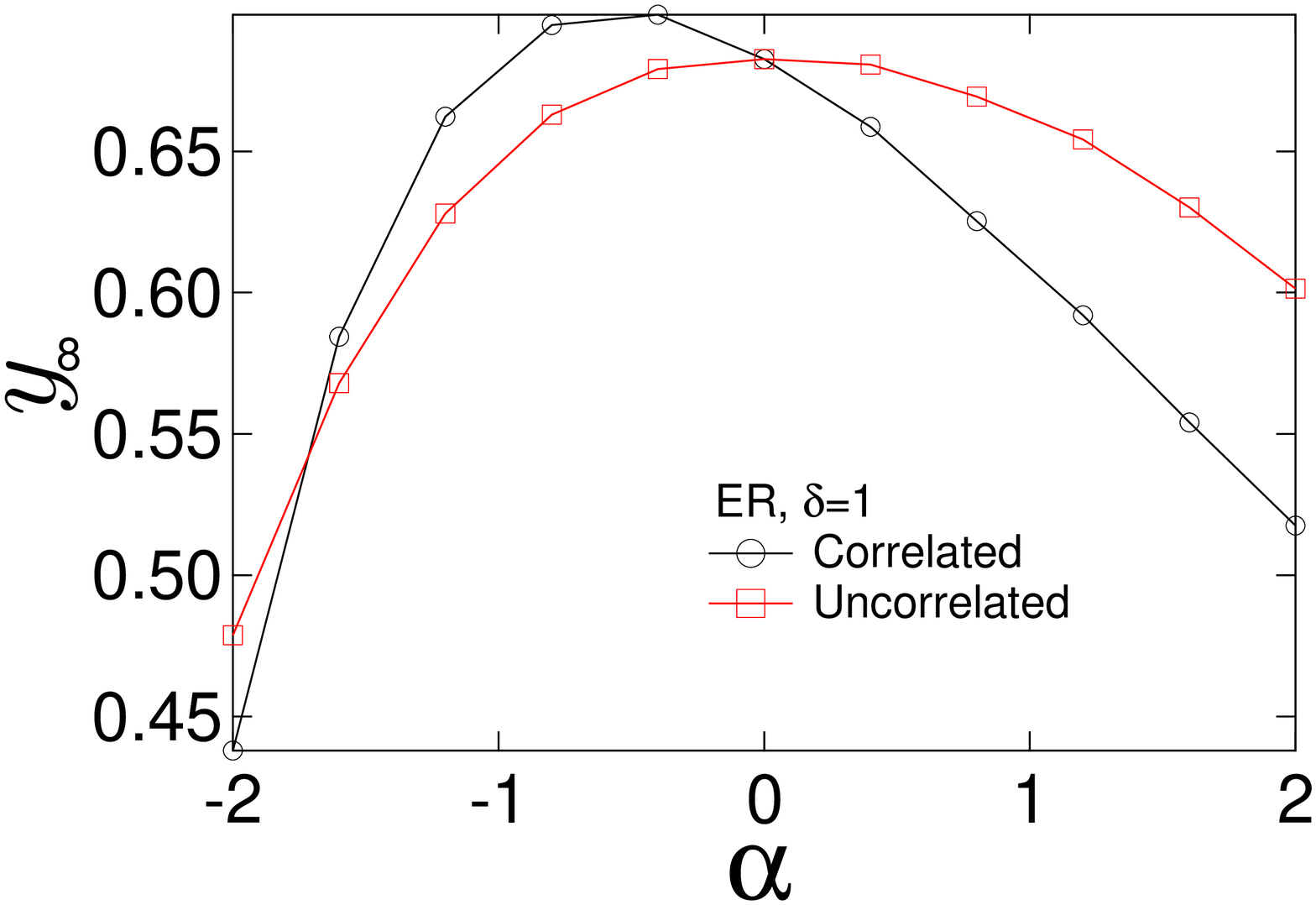}
\label{fig:14}
}
\subfigure[]{
\includegraphics[scale=.28]{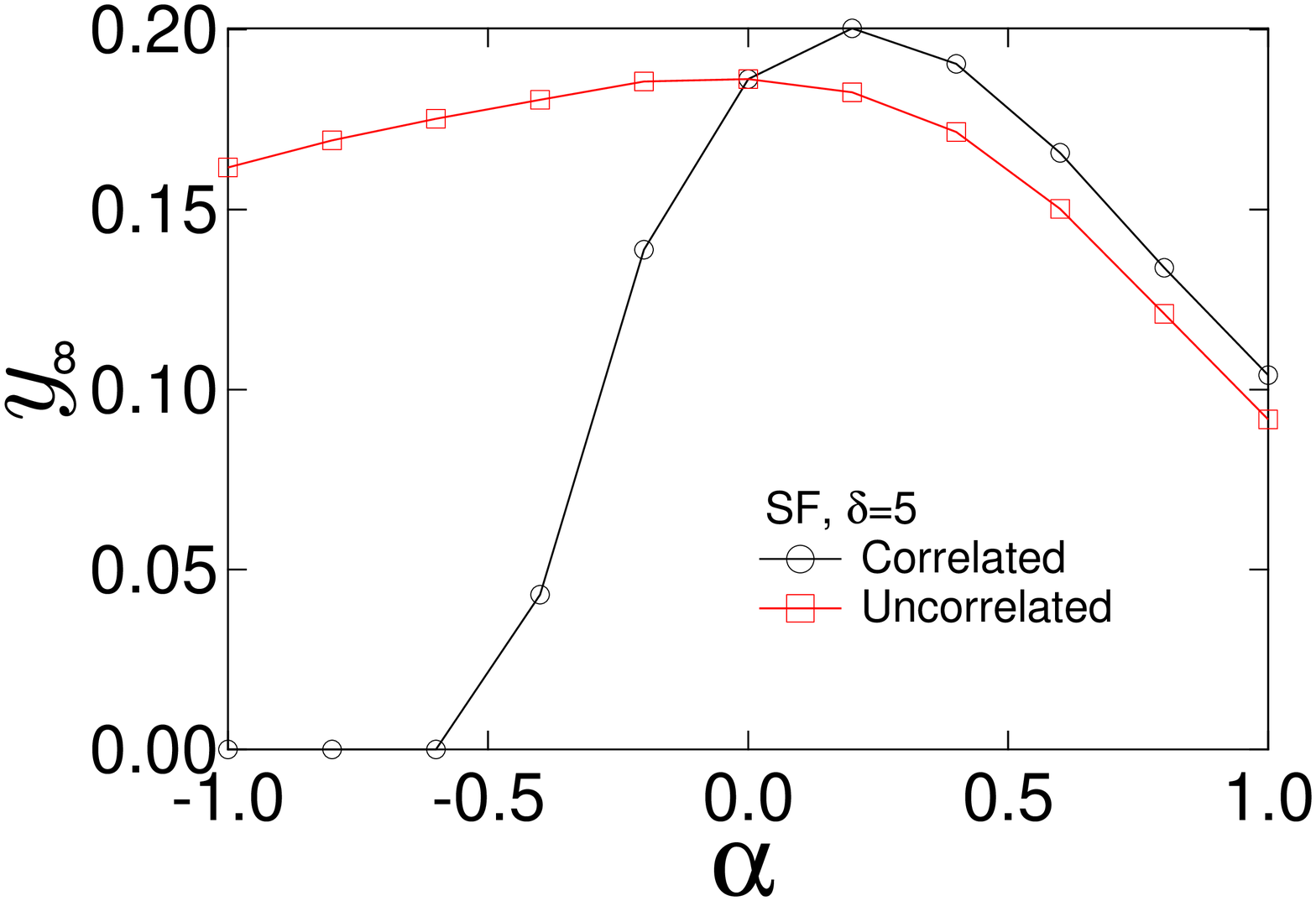}
\label{fig:15}
}
\subfigure[]{
\includegraphics[scale=.28]{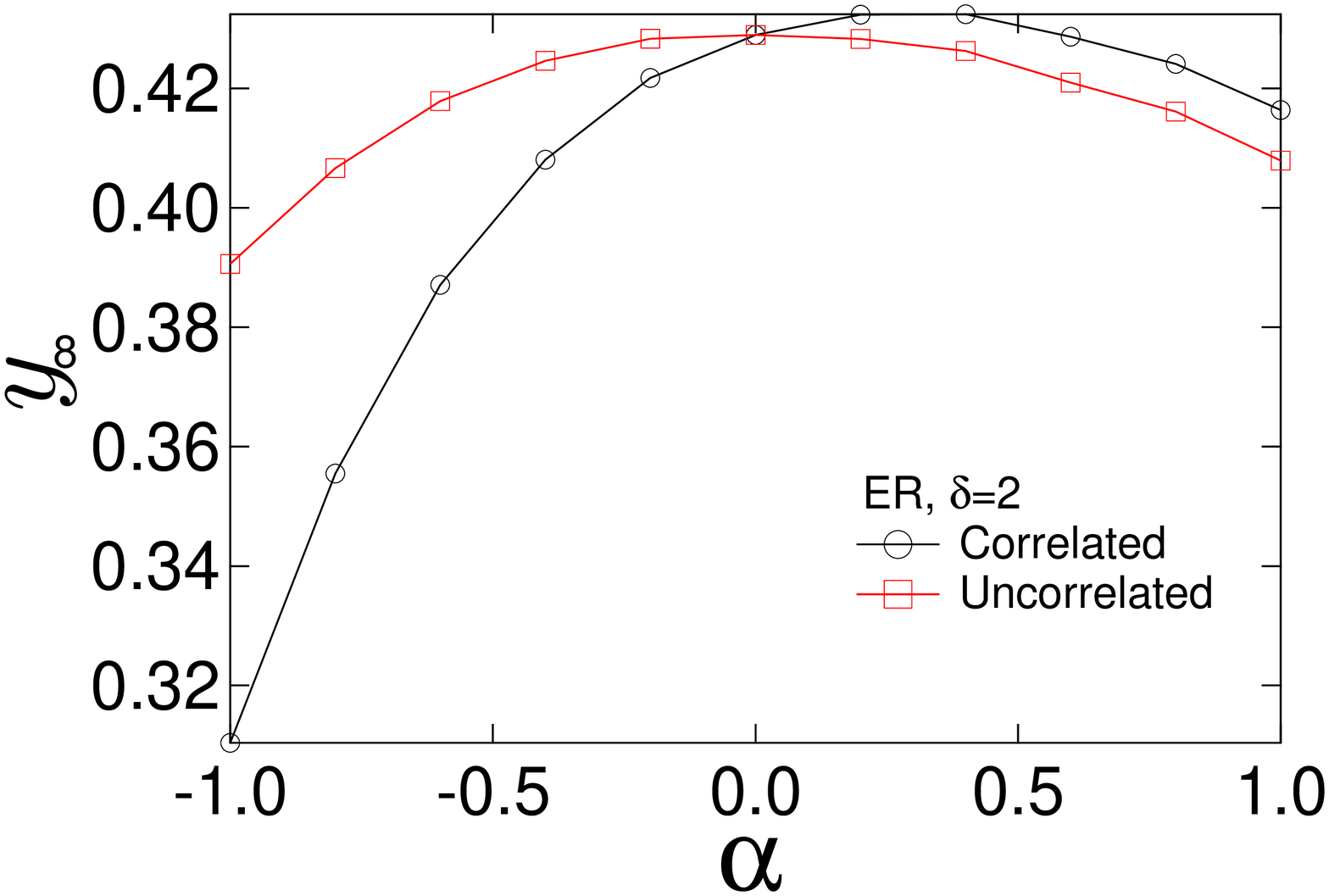}
\label{fig:16}
}

\caption{The average fraction $y_\infty$ as a function of $\alpha$ for (a) SF networks with the recovery rate $\delta=0.5$, (b) ER networks with the recovery rate $\delta=0.5$, (c) SF networks with the recovery rate $\delta=2$, (d) ER networks with the recovery rate $\delta=1$, (e) SF networks with the recovery rate $\delta=5$ and (f) ER networks with the recovery rate $\delta=2$ in both scenarios of correlated ($\circ$) and uncorrelated ($\square$) infection rates.}
\label{fig:10}
\end{figure}

The average fraction $y_\infty$ of infected nodes as a function of the scale parameter $\alpha$ in both ER and SF networks are shown in Fig.~\ref{fig:10}. We employ different values of the recovery rate $\delta$ to illustrate the influence of the correlation on the epidemic spreading over different range of recovery rates, i.e.\ different prevalence of the epidemic. 

Our previous work \cite{qu2015sis} explored the SIS model with i.i.d.\ heterogeneous infection rates and showed that the average fraction $y_\infty$ of infected nodes tends to decrease as the variance of the i.i.d.\ infection rates increases. In this paper, the uncorrelated infection rates can be considered as being independent. In this case, our previous results can be applied to derive the relationship between the average fraction $y_\infty$ of infected nodes and the parameter $\alpha$ as shown in Fig.~\ref{fig:10}: as $\alpha$ increases when $\alpha>0$ or as $\alpha$ decreases when $\alpha<0$, the variance of the infection rates increases, hence $y_{\infty}$ decreases in the scenario of uncorrelated infection rates. In the scenario of correlated infection rates, though the peaks of $y_\infty$ are not at $\alpha=0$ for both types of networks, $y_\infty$ roughly decreases as the absolute value of $\alpha$ increases. This is because, as the absolute value of $\alpha$ increases, large infection rates are assigned to a small number of links, limiting the spread of the epidemic.

From Fig.~\ref{fig:10}, we find that 1) the negative correlation ($\alpha<0$) between the infection rates and the degrees tends to enhance the epidemic spreading compared to the uncorrelated infection-rate scenario ($y_{\infty,c} > y_{\infty,u}$) when the recovery is small, but prohibit the spreading ($y_{\infty,c} < y_{\infty,u}$) when the recovery rate is large; 2) the positive correlation ($\alpha>0$) tends to enhance the epidemic spreading ($y_{\infty,c} > y_{\infty,u}$) when the recovery is large, but prohibit the spreading ($y_{\infty,c} < y_{\infty,u}$) when the recovery rate is small.

In the scenario of uncorrelated infection rates, the parameter $\alpha$ determines only the distribution of the i.i.d.\ infection rates in a network, and the infection rate between any pair of nodes is independent from their degrees. Compared to the scenario of uncorrelated infection rates, a positive correlation between the infection rates and nodal degrees ensures that the infection rates between nodes with larger degrees are also larger, whereas a negative correlation suggests the other way around. Intuitively, we may think that if the nodes with larger degrees can be infected by larger infection rates, the infection probability of those nodes are higher and those nodes can more effectively infect their neighbors as well. Hence, the positive correlation between the infection rates and nodal degrees seems to contribute to the epidemic spreading. This is indeed the case when the recovery rate $\delta$ is large, i.e.~the prevalence of the epidemic is low.

In contrast to this intuition, we observe that the positive correlation actually tends to prohibit the epidemic spreading when the recovery rate $\delta$ is small. This can be explained as follows: when the recovery rate $\delta$ is small, i.e.\ the prevalence is high, the infection probabilities of the large-degree nodes are already high, then the increment of the infection rates between the large-degree nodes may not significantly increase the infection probabilities of these nodes, and thus the infection probabilities of their neighbors may not be significantly increased. However, the negative correlation between the infection rates and nodal degrees leads to the higher infection rates between the small-degree nodes and effectively enhances the probabilities of the small-degree nodes compared to the scenario of the uncorrelated infection rates. Though the infection probabilities of large-degree nodes decrease in this case, the large amount of small-degree nodes ensures that the overall infection is on average enhanced.

\begin{figure}
\centering
\subfigure[]{
\includegraphics[scale=.28]{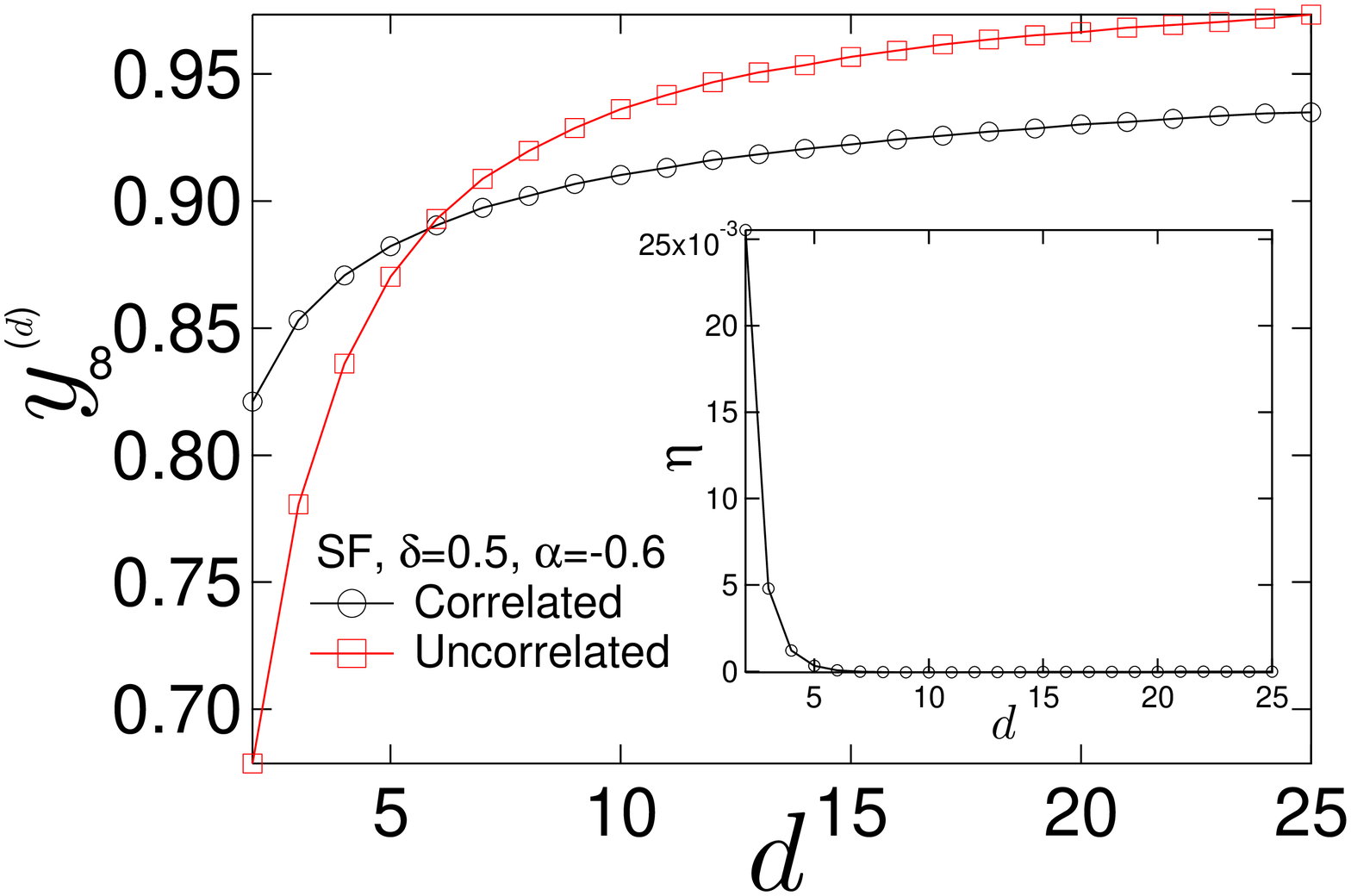}
\label{fig:21}
}
\subfigure[]{
\includegraphics[scale=.28]{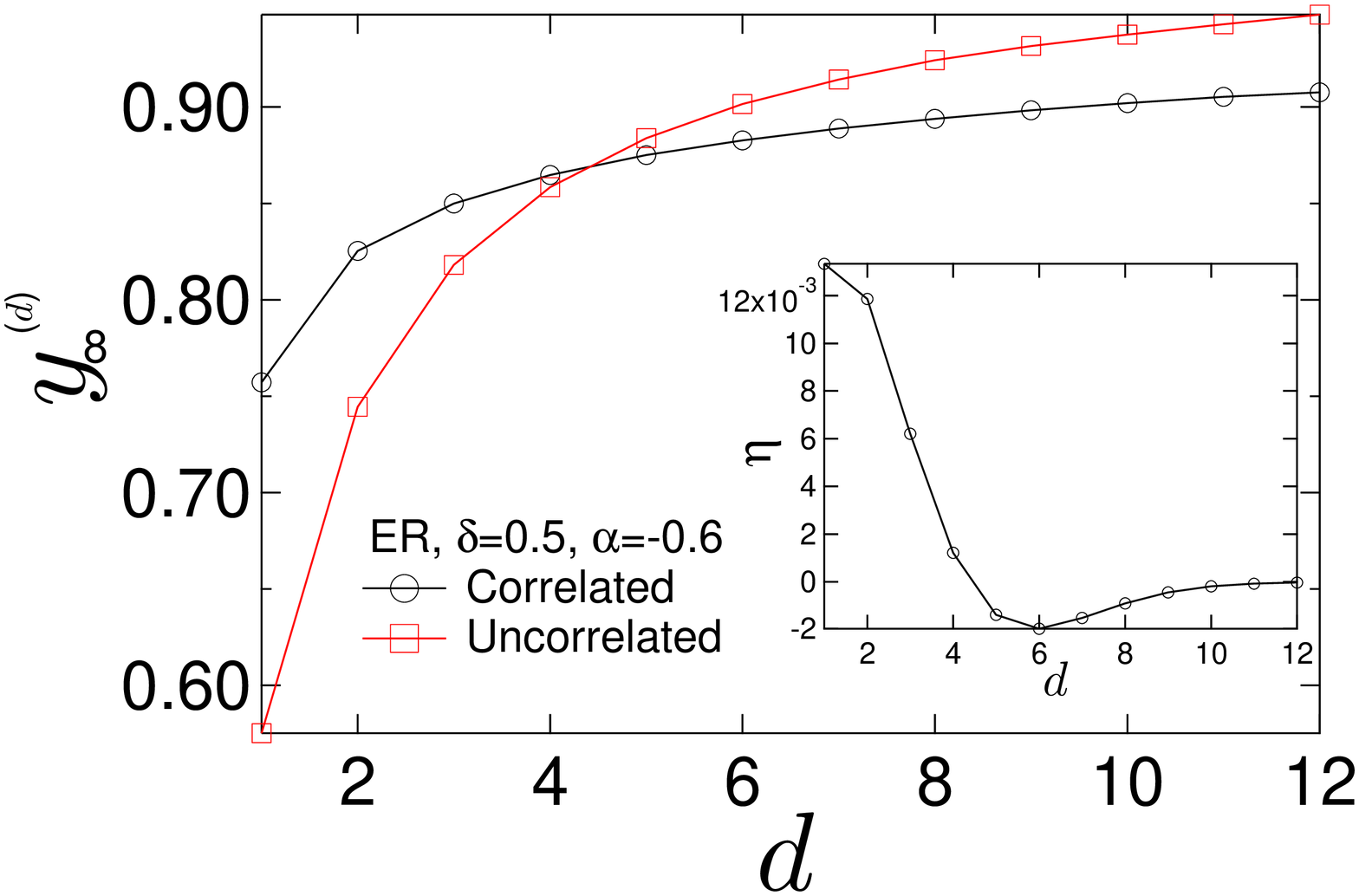}
\label{fig:22}
}
\caption{In the main figures: the average infection probability $y^{(d)}_\infty$ of the nodes with degree $d$ as a function of the degree $d$ for (a) SF networks and (b) ER networks with the recovery rate $\delta=2$ and $\alpha=-0.6$ for both scenarios of the correlated and uncorrelated infection rates. In the insets: $\eta$ as a function of $d$ with the same setting as the main figure.}
\label{fig:20}
\end{figure}

\begin{figure}
\centering
\subfigure[]{
\includegraphics[scale=.28]{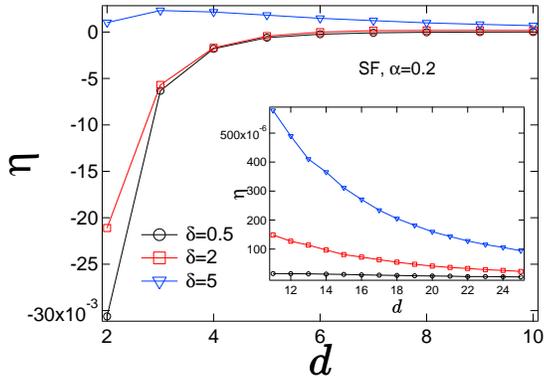}
\label{fig:31}
}
\subfigure[]{
\includegraphics[scale=.28]{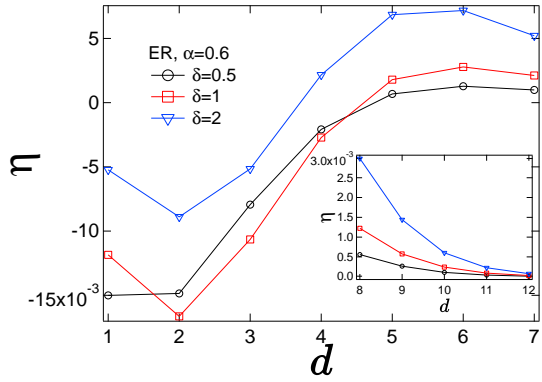}
\label{fig:32}
}
\subfigure[]{
\includegraphics[scale=.28]{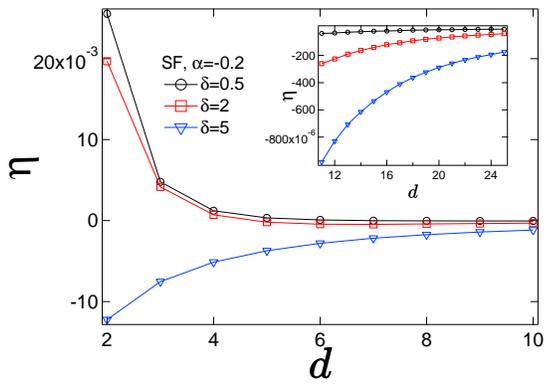}
\label{fig:33}
}
\subfigure[]{
\includegraphics[scale=.28]{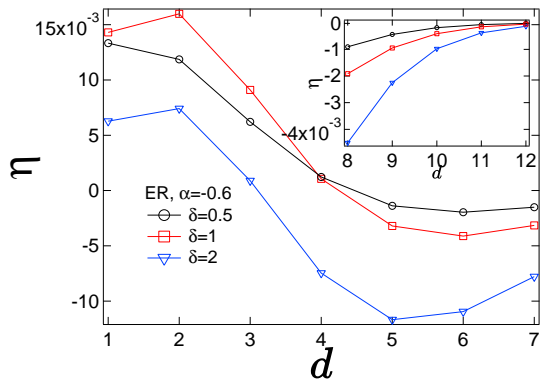}
\label{fig:34}
}
\caption{The plot of $\eta$ as a function of $\alpha$ for (a) SF networks and (b) ER networks with different recovery rates.}
\label{fig:30}
\end{figure}

To support our explanations, we define $y_\infty^{(d)}$ as the average infection probability of the nodes with degree $d$, and $y_{\infty,c}^{(d)}$ and $y_{\infty,u}^{(d)}$ as that for the scenario of correlated infection rates and uncorrelated infection rates respectively. We show $y_\infty^{(d)}$ as a function of the degree $d$ when the recovery rate is small, $\delta=0.5$, and the correlation parameter $\alpha=-0.6$ in Fig.~\ref{fig:20} (the main figures). We find that such a negative correlation ($\alpha=-0.6$) between the infection rates and nodal degrees indeed decreases the infection probabilities of large-degree nodes, but the infection probabilities of the small-degree nodes are also significantly lifted. Furthermore, the number of small-degree nodes is much larger than that of large-degree nodes in SF networks, and those are similar in ER networks. To illustrate the combined effect above of the two aspects, we define $\eta(d)$ as (\ref{equ:eta}), the product of the probability that a node has the degree $d$ and the difference between the average infection probability of the nodes with the degree $d$ in the scenarios of correlated and uncorrelated infection rates: 
\begin{equation}
\label{equ:eta}
\eta(d)=(y_{\infty,c}^{(d)}-y_{\infty,u}^{(d)})\text{Pr}[D=d]
\end{equation}
and $y_{\infty,c}-y_{\infty,u}=\sum_{d=1}^{d=N-1}\eta(d)$. Note that a positive $\eta(d)$ always indicates that $y_{\infty,c}>y_{\infty,u}$, i.e.\ the correlation lifts the infection probability of nodes with the degree $d$ compared to the scenario of the uncorrelated infection rates. As in the insets of Fig.~\ref{fig:20}, we plot $\eta(d)$ as a function of the degree $d$ for both ER and SF networks. 
More plots\footnote{We still select $\alpha=-0.6$ as the example of the negative correlation and $\alpha=0.6$ as the comparison for ER network, but we select $\alpha=-0.2$ as the example of the negative correlation (and $\alpha=0.2$ as the comparison) for SF networks since when $\delta=5$ and $\alpha=-0.6$ the epidemic already dies out in the scenario of correlated infection rates.} of $\eta$ as a function of the degree $d$ are shown in Fig.~\ref{fig:30} where the cases with small degrees are shown in the main figures and those with relatively large degrees are shown in the insets.
We find that in both networks the value of $\eta$ is significantly large for the small-degree nodes and contributes more to a higher prevalence of the epidemic when the recovery rate is small and the correlation is negative as shown in Fig.~\ref{fig:20}, Fig.~\ref{fig:33} and Fig.~\ref{fig:34}.
The observation is consistent with our explanation about why the negative correlation tends to help the epidemic spreading when the recovery rate is small. In contrast, the observation, shown in Fig.~\ref{fig:31} and Fig.~\ref{fig:32}, that the positive correlation does increase the infection probabilities of large-degree nodes but decreases those of small-degree nodes more when the recovery rate is small, also supports our explanation about how the positive correlation prohibits the spreading when recovery rate is small.  

As the recovery rate becomes large thus the prevalence is low, the positive but not negative correlation, between the infection rates and nodal degrees may effectively enhance the infection probabilities of the large degree nodes comparing to the uncorrelated infection rates. 
When the recovery rate $\delta$ is large, for example, $\delta=5$ for SF networks and $\delta=2$ for ER networks in this paper, the positive correlation leads to the increment of the infection probabilities of large-degree nodes and could further lift the probabilities of their small-degree neighbors which are large in number. The infection probabilities of small-degree nodes are reduced, but the infection probabilities are already low and the small-degree nodes have few neighbor to infect. Hence, the overall infection increases on average. 
The explanations are supported by Fig.\ \ref{fig:31} and Fig.\ \ref{fig:32}. The infection probabilities of the large-degree nodes increases for different recovery rates when $\alpha$ is positive, and the increment of the infection probabilities of large-degree nodes is also larger as the recovery rate increases. Thought the infection probabilities of the small-degree nodes may decrease when the correlation is negative, the increment of the infection probabilities of the large-degree nodes also lifts the infection probabilities of the small-degree nodes. As a result, the infection probabilities of majority nodes are on average lifted when the recovery rate $\delta=5$  and $\delta=2$ for SF and ER networks respectively.

%In contrast, though the negative correlation may increase the infection probabilities of the small-degree nodes, the decrement of the infection probability of the large-degree node tends to lower down the infection probability of the small-degree nodes, so the infection probability of the small-degree nodes may not significantly increase or even decrease when the recovery rate is large. Hence, the overall infection probability decreases when the correlation is negative and the recovery rate is large. This conclusion is supported by Fig.\ \ref{fig:33} and Fig.\ \ref{fig:34}. The infection probabilities of the large-degree nodes are always lowered down under the negative correlation, and as the recovery rate increases, the infection probabilities of the majority nodes for both SF and ER networks also tends to decreases. 

For both ER and SF networks, the positive correlation between the nodal degrees and the infection rates tends to enhance the spreading when the recovery rate is small, whereas the negative correlation tends to help when the recovery rate is large.
As the recovery rate $\delta$ increases from $0$ and if the absolute value of $\alpha$ is small, we expect that there is a critical value $\delta_c$: when $\delta<\delta_c$ the negative correlation tends to enhance the spreading, otherwise ($\delta>\delta_c$) the positive correlation is likely to help the spreading. By the comparing Fig.~\ref{fig:13} and Fig.~\ref{fig:16}, we can observe that $\delta_c$ is larger in SF networks than ER networks. This difference is mainly caused by that the prevalence in SF networks tends to be higher than that in ER networks when the recovery rate and the parameter $\alpha$ are the same and the positive correlation tends to enhance the epidemic spreading when the prevalence is low as we discussed.

\subsection{Extreme cases}  
\label{Sec:largeA}
We then discuss the influence of the correlation between the infection rates and the nodal degrees when the correlation is strong, i.e.\ the absolute value of $\alpha$ is large. 

When the absolute value of $\alpha$ is large, the variance of the infection rates is large as well. In this case, most links possess a small infection rate and few have a large infection rate. A large proportion of the links have such a small infection rate that the infection processes driven by the small infection rate will hardly happen. The networks are actually filtered by the small infection rates. The other small proportion of the large infection rates then mainly determine the overall infection. In the scenario of the uncorrelated infection rates, the few large infection rates are randomly distributed, thus cannot form a connected cluster. Compared to the scenario of the uncorrelated infection rates, the positive correlation ensures that the large infection rates are distributed between the large-degree nodes which are more likely to connect with each other, forming a subgraph. A connected subgraph tends to help the spread of an epidemic. In the other way around, the negative correlation will almost surely stop the epidemic spreading since the few large infection rates distributed between the small-degree nodes could hardly form such a connected cluster. In summary, the positive correlation enhances the epidemic spreading whereas the negative one prohibits when the absolute value of $\alpha$ is large. 

\section{The wheel network}
\label{Sec_wheel}
In this section, we further consider a special topology -- the wheel network. We are going to prove that, compared to the uncorrelated heterogeneous infection rate, the negative correlation between the infection rates and the degrees tends to help the epidemic spreading in a large wheel network when the recovery rate is small, whereas the positive correlation tends to contribute to the epidemic spreading when the recovery rate is large. 

In a wheel network, $m$ side nodes compose a ring, i.e.\ node $i$ connects with node $i+1$ ($i=1,2,3,...,m-1$) and node $m$ connects with node $1$, and all the $m$ side nodes connect with one central node -- node $0$. In this section, we consider a large enough wheel network, and without loss of the generality, we still set both the homogeneous infection rate and the average of heterogeneous infection rates to be $1$, and tune the recovery rate. 

In the scenario of the uncorrelated infection rates, the infection rates are actually i.i.d. In our previous work, we found that, compared to the homogeneous infection rate, the i.i.d.\ heterogeneous infection rates always reduce the overall infection if the epidemic can spread out. We further verify this conclusion for the wheel network by simulations as shown in \ref{App:Wheel}, where we plot the average fraction of infection nodes as a function of $\alpha$ for the scenario of uncorrelated infection rates and different values of the recovery rates. 
That is to say, the average fraction of infected nodes reaches the maximum when the infection rates are homogeneous, i.e.\ $\alpha=0$ in the scenario of the uncorrelated infection rates. If the average fraction of infected nodes in the scenario of the correlated infection rates is larger than that when the infection rates are homogeneous, then the correlation enhances the epidemic spreading compared to the uncorrelated case. 

We first consider the homogeneous infection rate in a wheel network, where the infection rates are homogeneous thus the same for all links. The infection probability $v_{0\infty}$ of the central node is 
\begin{equation}
\label{equ:homo-wheel-central}
v_{0\infty}=1-\frac{1}{1+m\tau v_{i\infty}}=1-\frac{\delta}{\delta+mv_{i\infty}}
\end{equation}
where $v_{i\infty}$ is the infection probability of a side node, which is the same for all the side nodes. The infection probability of the side node is 
\begin{equation}
\label{equ:homo-wheel-side}
v_{i\infty}=1-\frac{1}{1+2\tau v_{i\infty}+\tau v_{0\infty}}
=1-\frac{\delta}{\delta+2 v_{i\infty}+ v_{0\infty}}
\end{equation}
By solving (\ref{equ:homo-wheel-central}) and (\ref{equ:homo-wheel-side}), we obtain the positive solution 
\begin{equation}
\label{equ:homo-wheel-central-root}
v_{0\infty}=\frac{m-\delta^2+2\delta+\frac{\delta \left((\delta-1)m+2\delta+\sqrt{(\delta^2-2\delta+9)m^2-(4\delta^2+12\delta)m+4\delta^2}\right)}{2m}}{\delta+m}
\end{equation}
and
\begin{equation}
\label{equ:homo-wheel-side-root}
v_{i\infty}=\frac{(1-\delta)m-2\delta+\sqrt{(\delta^2-2\delta+9)m^2-(4\delta^2+12\delta)m+4\delta^2}}{4m}
\end{equation}
We consider a large $m$ and a constant $\delta$. In this case, $v_{0\infty}\approx 1$ and 
\begin{equation}
\label{equ:homo-wheel-small-y}
 v_{i\infty}\approx\frac{1-\delta+\sqrt{\delta^2-2\delta+9}}{4}
\end{equation}
The fraction of infected nodes is then
\[
y_\infty=\frac{mv_{i\infty}+v_{0\infty}}{m+1}\approx v_{i\infty}
\]

Now we consider the wheel network where the infection rates are correlated as defined before in (\ref{eq1}). There are two kinds of infection rates in a wheel network: 1) the infection rate $\beta_0$ between the central node and a side node, i.e.\ $\beta_0\sim (3m)^{\alpha}$; 2) the infection rate $\beta_1$ between a pair of connected side nodes, i.e.\ $\beta_1\sim (3*3)^{\alpha}$. Since we consider a large $m$, we find $\beta_0\gg \beta_1$ if $\alpha> 0$, whereas $\beta_0\ll \beta_1$ if $\alpha< 0$. When the correlation between the infection rates and the degree is positive, the infection rates $\beta_0\approx 2$ and $\beta_1\approx 0$ and the network becomes a star network. In contrast, when the correlation is negative, the infection rates $\beta_1\approx 2$ and $\beta_0\approx 0$ and the network becomes a ring. By NIMFA, we can compute the average fraction of infected nodes\footnote{Equation (\ref{equ:star}) can be similarly derived by solving (\ref{Equ:NIMFA-Steady}).} when $\alpha>0$ 
\begin{equation}
\label{equ:star}
y_\infty\approx v_{i\infty}=\frac{2}{2+\delta}
\end{equation}
When $\alpha<0$, the average fraction of infected nodes\footnote{Equation (\ref{equ:ring}) can be derived by applying the Laurent series as in (\ref{Equ:NIMFA-series}).} 
\begin{equation}
\label{equ:ring}
y_\infty=v_{i\infty}=1-\frac{\delta}{4}
\end{equation}
The parameter $\alpha=0$ in the scenario of the correlated infection rates indicates the homogeneous infection rate which is the same as $\alpha=0$ in the scenario of the uncorrelated infection rates. We have shown that the average fraction of infected nodes reaches the maximum when $\alpha=0$ in the scenario of the uncorrelated infection rates, so we further compare the average fraction of infected nodes as shown in (\ref{equ:homo-wheel-small-y}) when the infection rates are homogeneous i.e.\ $\alpha=0$ with that as shown in (\ref{equ:ring}) when $\alpha<0$ or that as shown in (\ref{equ:star}) when $\alpha>0$ to explore the influence of the correlation on the epidemic spreading. By comparing (\ref{equ:homo-wheel-small-y}) and (\ref{equ:ring}), we find that when the recovery rate $\delta<2$, the average fraction $y_\infty$ of infected nodes is higher if $\alpha<0$ than if $\alpha=0$. Hence, the negative correlation between the infection rates and nodal degrees helps the epidemic spreading if the recovery rate is small, i.e.\ $\delta<2$. Similarly, if we compare (\ref{equ:homo-wheel-small-y}) and (\ref{equ:star}), we find that when the recovery rate $\delta>2$ the average fraction of infected nodes is higher if $\alpha>0$ than if $\alpha=0$, and conclude that the positive correlation between the infection rates and nodal degrees contributes to the epidemic spreading if the recovery rate is large. The theoretical results are consistent with our previous conclusions: when the recovery rate is small, the negative correlation tends to help the epidemic spreading, and as the recovery rate increases to be larger than a critical value, i.e.\ $2$ in this case, the positive correlation enhances the spreading. 

\section{Real-world networks}     
As mentioned, the interaction frequency $\beta_{ij}$ between node $i$ and node $j$ in a real-world network can be considered as the infection rate between them and it has been found that $\beta_{ij}\sim (d_id_j)^{\alpha}$ in many networks. In this section, we choose two real-world networks as examples to illustrate how their heterogeneous infection rates affect the spread of SIS epidemics on these networks. We compare the average fraction $y_\infty$ of infected nodes in the metastable state of the two networks in the two scenario: 1) the scenario of correlated infection rates, where each network is equipped with its original heterogeneous infection rates (but normalized so that the average infection rate is $1$) as given in the dataset; 2) the scenario of uncorrelated infection rates, where each network is equipped with the infection rates (normalized as well) in the original dataset but the infection rates are shuffled and reassigned to each link. Our objective is to explore the relation between the infection rates and average fraction of infected nodes in these $2$ scenarios for both networks to verify our previous findings. 

The first network is the airline network (with $3071$ nodes and $15358$ links) where the nodes are the airports and the infection rate along a link is the number of flights between the two airports. We construct this network and its infection rates from the dataset of openFlights\footnote{http://openflights.org/data.html}. The other one is the co-author network (with $39577$ nodes and $175692$ links) where the nodes are the authors of papers, and the infection rate is the collaboration frequency depending on the number of collaborated papers and the number of authors in those papers\cite{newman2001structure}. 

\begin{figure}
\centering
\subfigure[]{
\includegraphics[scale=.28]{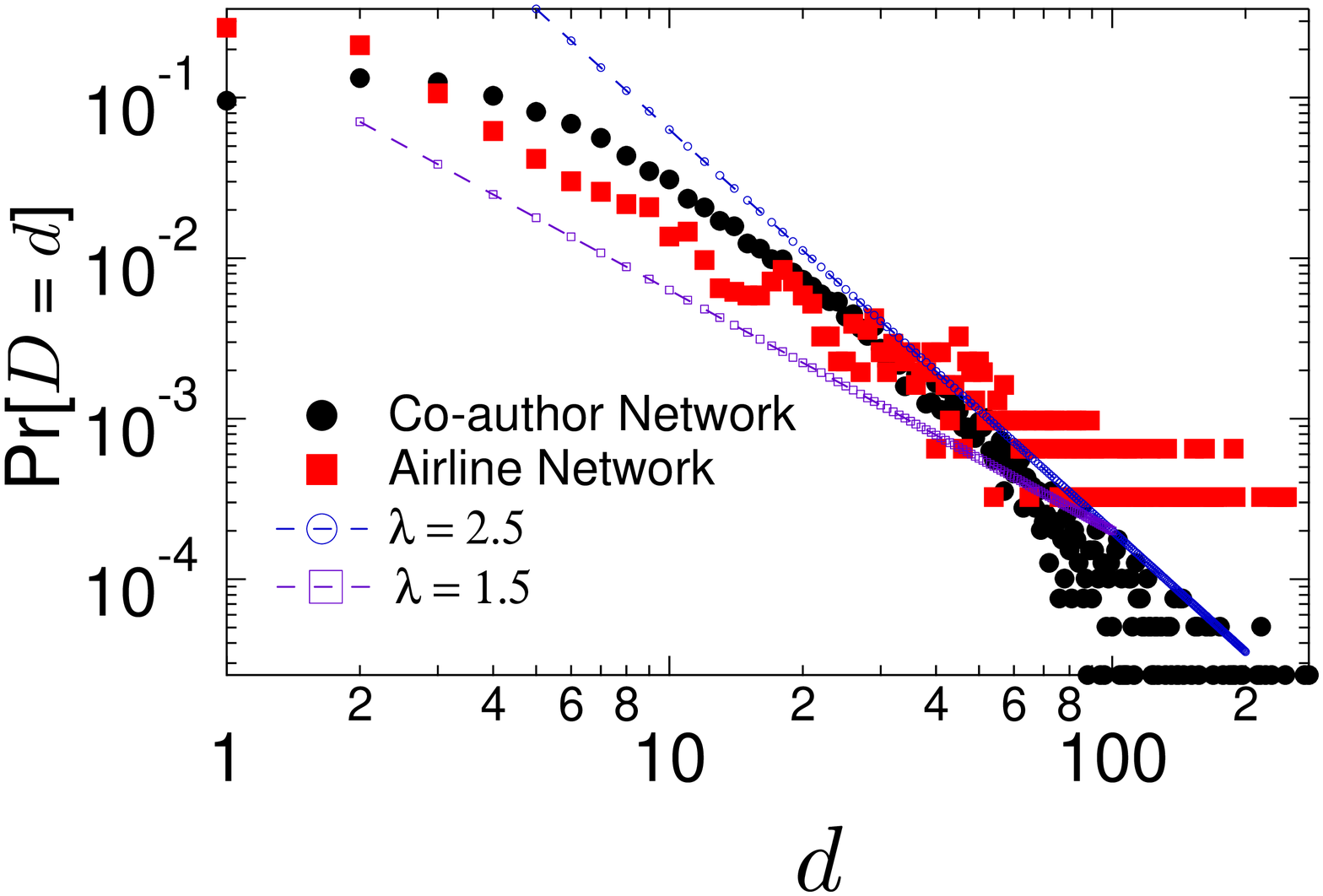}
\label{fig:51}
}
\subfigure[]{
\includegraphics[scale=.28]{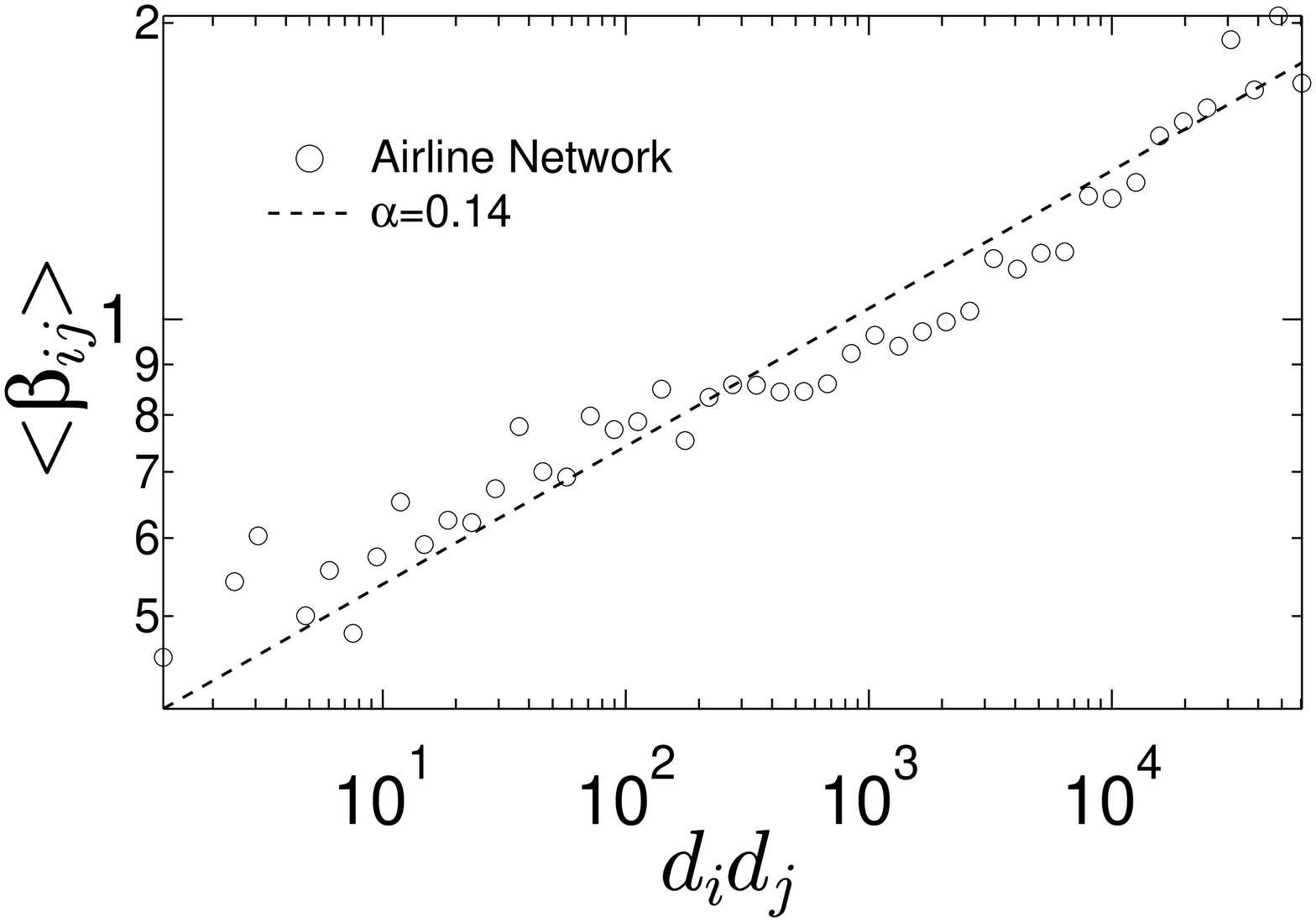}
\label{fig:52}
}
\subfigure[]{
\includegraphics[scale=.28]{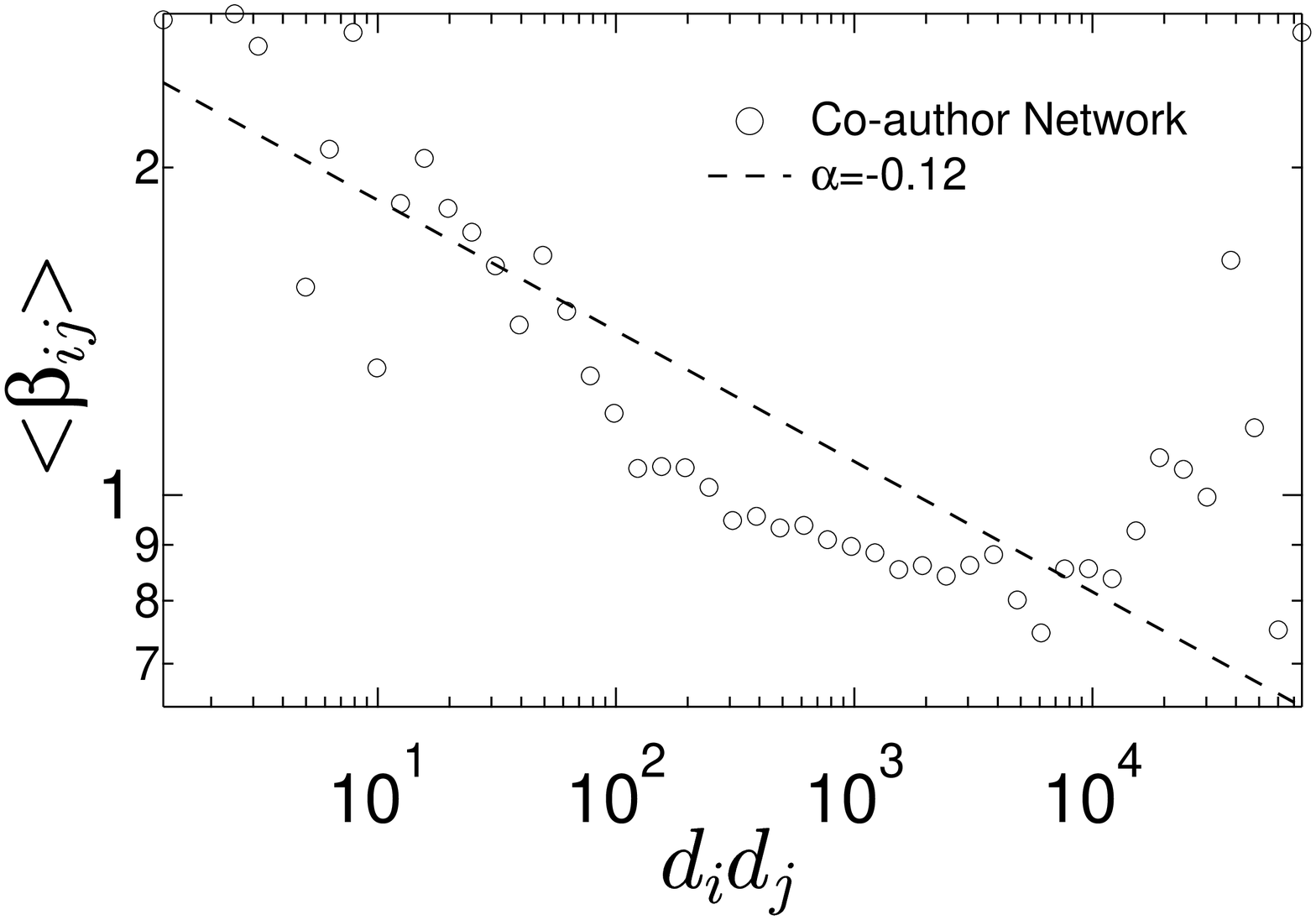}
\label{fig:53}
}
\caption{(a) The degree distributions of two real-world networks: the airline and co-author networks. (b) The interaction frequency $\beta_{ij}$ as a function of the nodal degrees $d_id_j$ for the airline network. (c) The interaction frequency $\beta_{ij}$ as a function of the nodal degrees $d_id_j$ for the co-author network.}
\label{fig:5}
\end{figure}

As shown in Fig.~\ref{fig:51}, the degree distributions of the airline network and co-author network approximately follow a power law with the slope $\lambda=1.5$ and $2.5$ respectively. Moreover, we plot the average infection rates $\langle\beta_{ij}\rangle$ as a function of the product of the two nodal degrees $d_id_j$ in Fig.~\ref{fig:52} and \ref{fig:53} for the airline and co-author networks respectively. We find that, roughly $\beta_{ij}\sim (d_id_j)^\alpha$ with $\alpha=0.14$ and $\alpha=-0.12$ for the airline and co-author networks respectively. 

\begin{figure}
\centering
\subfigure[]{
\includegraphics[scale=.28]{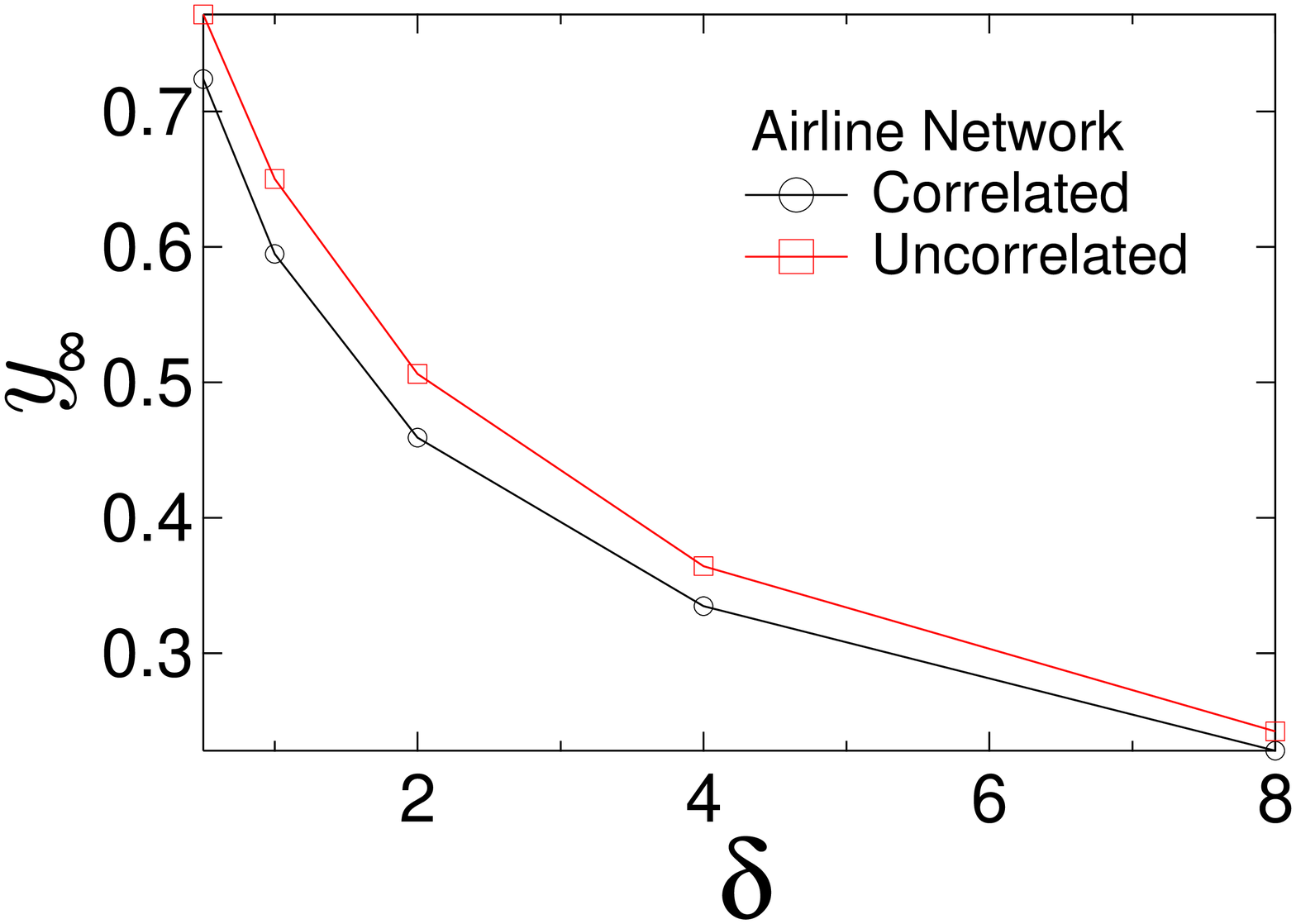}
\label{fig:61}
}
\subfigure[]{
\includegraphics[scale=.28]{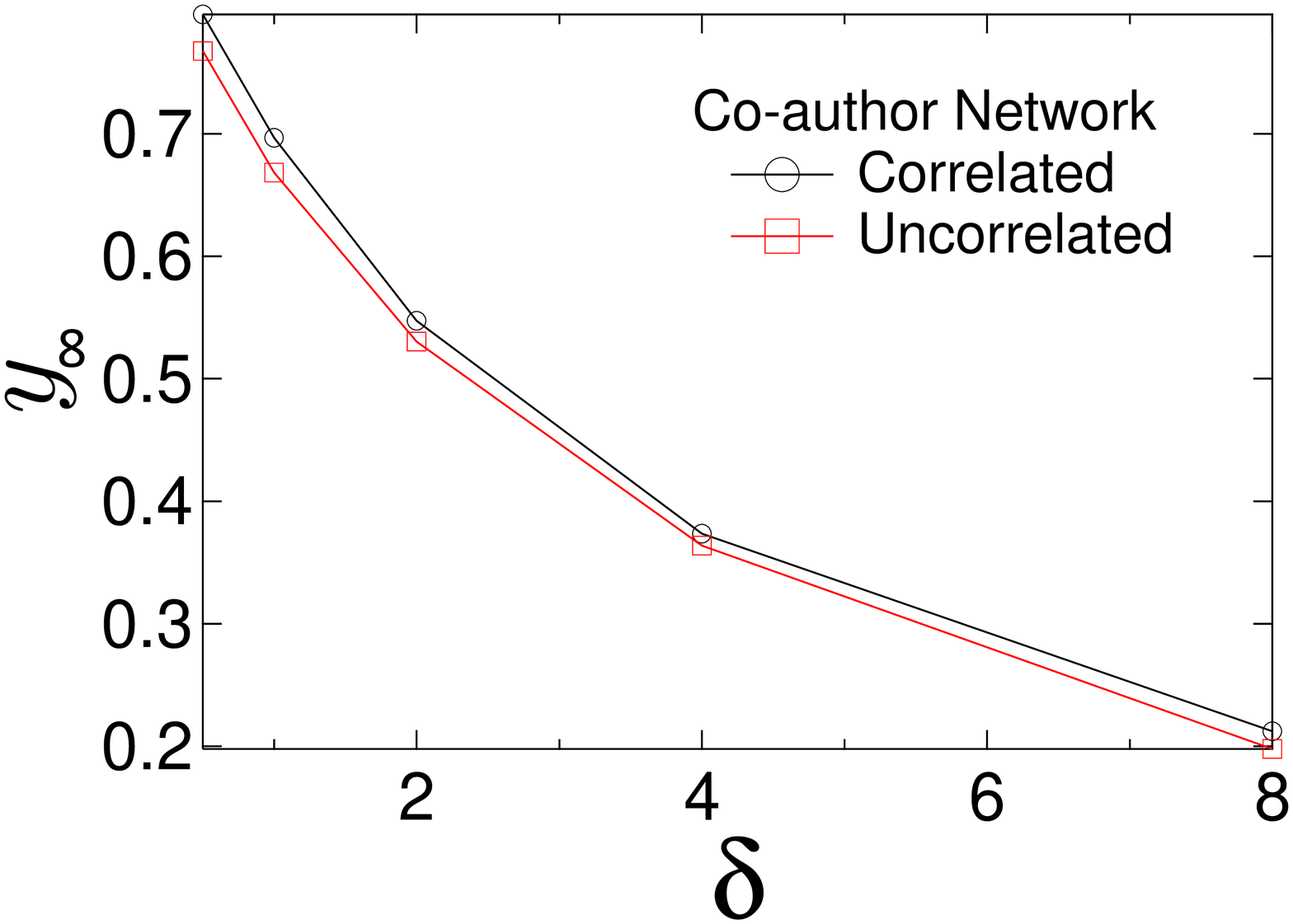}
\label{fig:62}
}
\caption{The average fraction $y_\infty$ as a function of the recovery rate $\delta$ for (a) the airline network and (b) the co-author network in both scenarios of correlated ($\circ$) and uncorrelated ($\square$) infection rates. The recovery rate is in the range $[0.5,8]$. }
\label{fig:6}
\end{figure}

\begin{figure}
\centering
\subfigure[]{
\includegraphics[scale=.28]{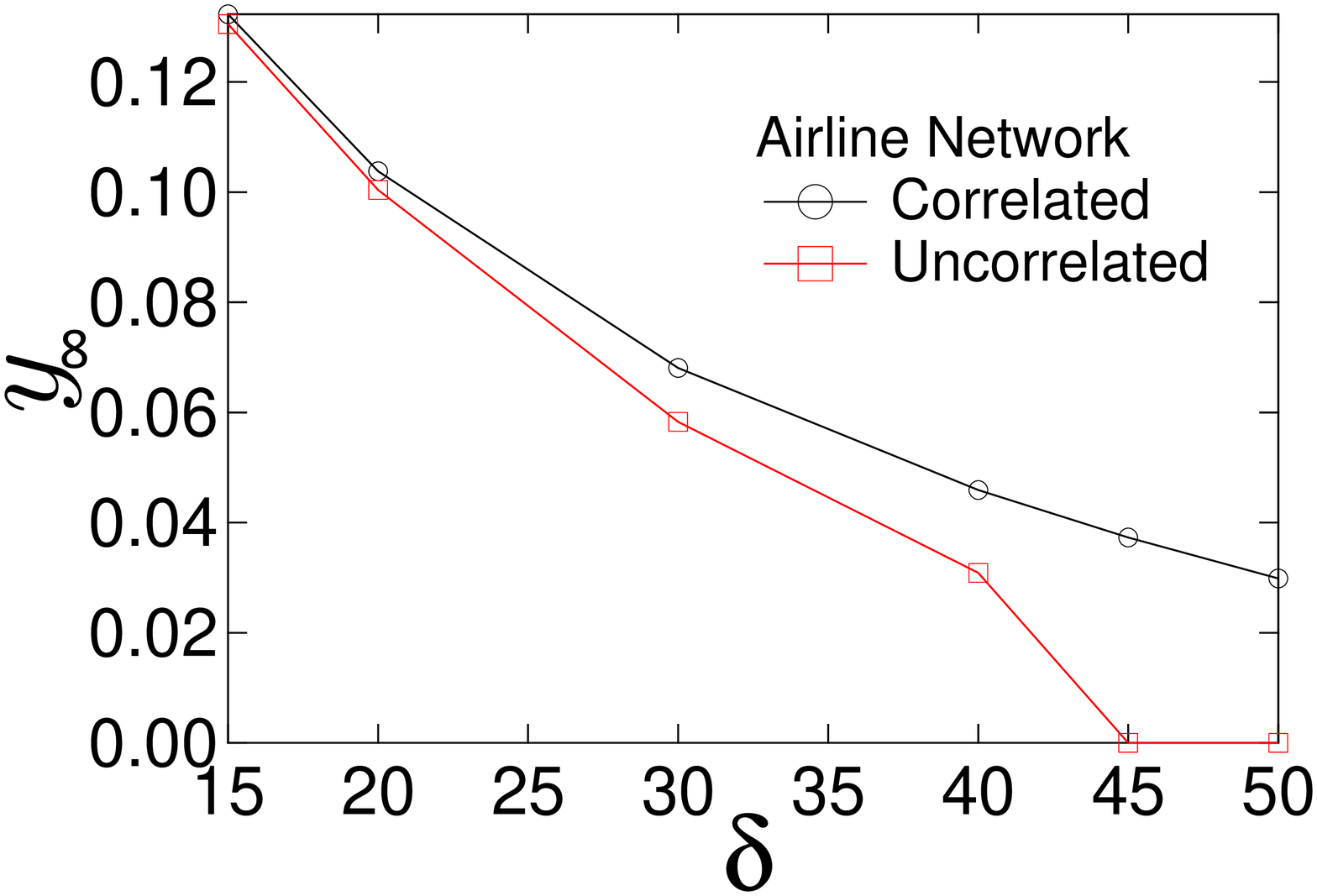}
\label{fig:71}
}
\subfigure[]{
\includegraphics[scale=.28]{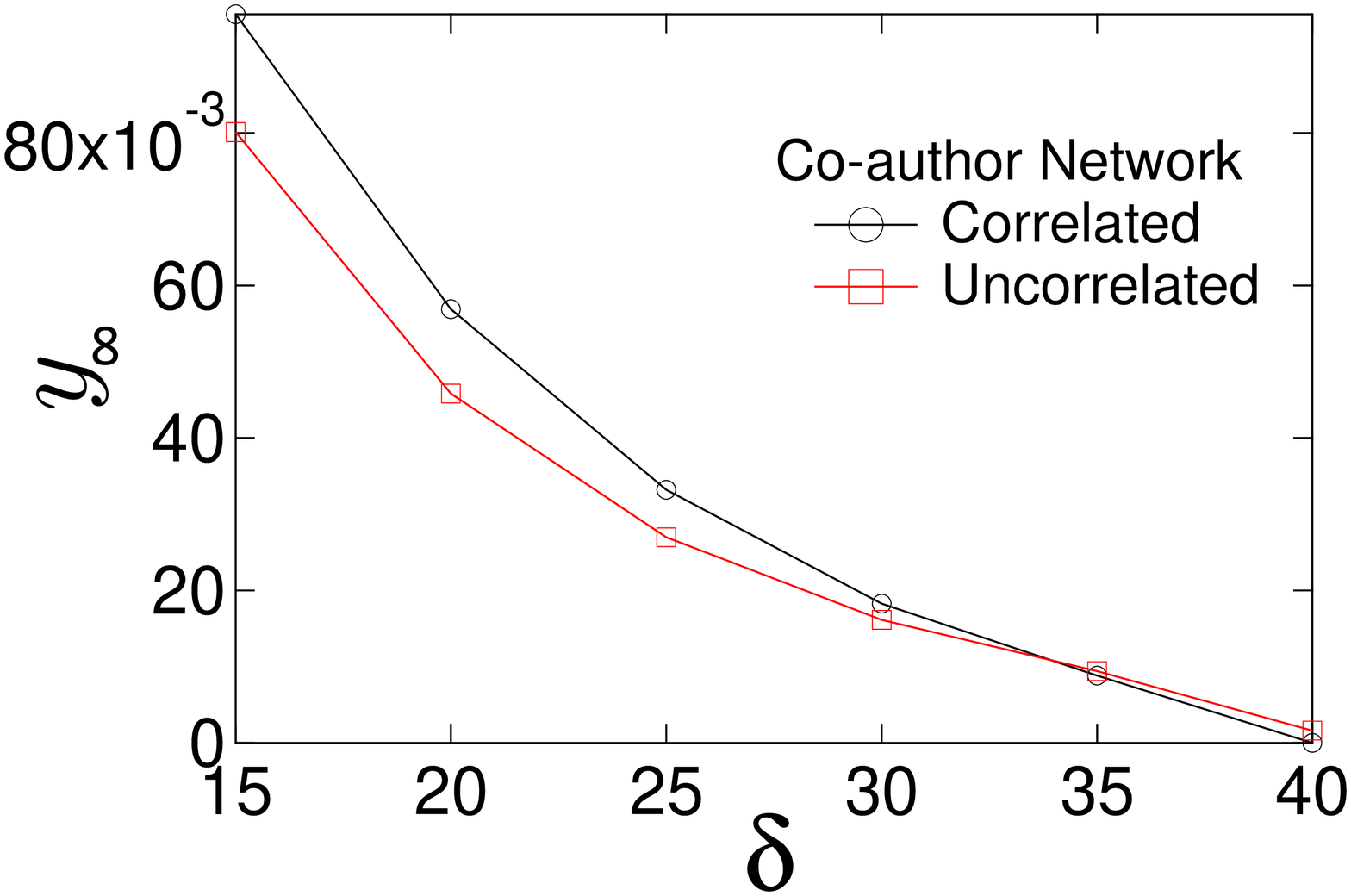}
\label{fig:72}
}
\caption{The average fraction $y_\infty$ as a function of the recovery rate $\delta$ for (a) the airline network and (b) the co-author network in both scenarios of correlated ($\circ$) and uncorrelated ($\square$) infection rates. The recovery rate is large.}
\label{fig:7}
\end{figure}

We first discuss the case when the recovery rate is small, i.e. $\delta\in[0.5,8]$ (as shown in Fig.~\ref{fig:6}), and then the case when $\delta$ is large, i.e. $\delta \ge15$ (as shown in Fig.~\ref{fig:7}), since for both networks the recovery rate $\delta\in[0.5,8]$ enables the high prevalence of the epidemic and $\delta \ge15$ leads to the low prevalence. The average fraction $y_\infty$ of infected nodes as a function of the recovery rate $\delta$ for both infection-rate scenarios is shown in in Fig.~\ref{fig:6} when the recovery rate $\delta$ is small. We find that the positive correlation ($\alpha=0.14$) between the infection rates and nodal degrees in the airline network retards the spread of epidemics, whereas the negative correlation ($\alpha=-0.12$) in the co-author network contributes in the other way around. 
The observations are consistent with our previous conclusions that the positive correlation between the infection rates and nodal degrees probably retards the spread of epidemics whereas a negative correlation may help the spread when the recovery rate is small. The average fraction $y_\infty$ of infected nodes as a function of the recovery rate $\delta$ when the recovery rate is large is shown in Fig.~\ref{fig:7}.
We find that the positive correlation $\alpha=0.14$ in the airline network (Fig.~\ref{fig:71}) helps the epidemic spreading when the recovery rate is larger than $15$, and the negative correlation $\alpha=-0.12$ in the co-author network suppresses the epidemic spreading only when the recovery rate is large, i.e.\ $\delta\le 35$, that the overall infection is close to $0$. The observations also agree with our conclusion that the positive correlation between the infection rates and nodal degrees tends to help the epidemic spreading but not the negative correlation when the recovery rate is large. Hence, the simulation results of the CSIS model on the real-world networks for both the small and large recovery rates agree with our previous conclusions.  

\section{Conclusion}
In this paper, we study how the correlation between the infection rates and nodal degrees influences the epidemic spreading, compared to the uncorrelated case. By continuous-time simulations of our CSIS model in different infection-rate scenarios and networks with different topology heterogeneities, we find that, when the recovery rate is small, i.e.\ the prevalence of the epidemic is high, the negative correlation between the nodal degree and the infection rates tends to help the epidemic spreading. However, when the prevalence is high, the positive correlation is more likely to enhance the spreading. The validation on two real-world networks and the proof in the large wheel network agree with our conclusions. Our results shed light on that how the epidemic spreads in the real-world could be far away from the simple classic models. Not only the heterogeneity of infection rates but also the correlation between the heterogeneity of infection rates and network topologies could be various and complicated. Our work is the first step to study the correlated heterogeneous SIS model in heterogeneous networks.

\appendix

\section{The wheel network with i.i.d.\ infection rates}
\label{App:Wheel}
\begin{figure}[h]
\centering
\includegraphics[scale=.4]{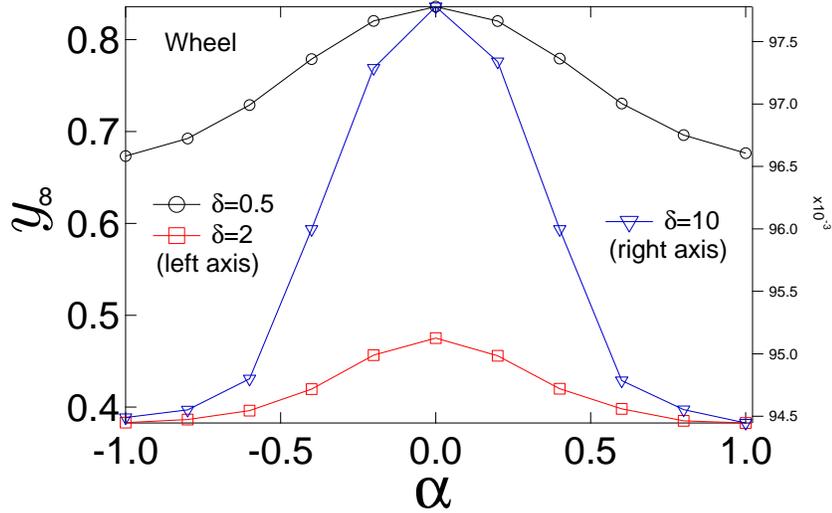}
\caption{The average fraction $y_\infty$ as a function of the correlation strength parameter $\alpha$ for a wheel network with $m=1000$ side nodes in the scenario of the uncorrelated infection rates. The recovery rate is $\delta=0.5$ ($\bigcirc$),  $\delta=2$ ($\square$) and $\delta=10$ ($\bigtriangledown$). }
\label{fig:A1}
\end{figure}

As shown in Fig.~\ref{fig:A1}, we plot the average fraction $y_\infty$ of infected nodes as a function of the correlation strength parameter $\alpha$ for the wheel network with $m=1000$ side nodes in the scenario of the uncorrelated heterogeneous infection rates. We employ the recovery rates $\delta=0.5$, $\delta=2$ and $\delta=10$. We find that no matter what value the recovery rate is, the homogeneous infection rate ($\alpha=0$) always leads to the highest overall infection.  

%% References

\end{document}